\pdfoutput=1
\documentclass[a4paper,12pt]{article}
\usepackage[utf8]{inputenc}
\usepackage{placeins}
\usepackage{amssymb,amsmath,amsfonts}
\usepackage[colorlinks,linkcolor=purple,citecolor=teal]{hyperref}
\usepackage{dsfont}
\usepackage{color}
\usepackage{graphicx}
\usepackage{hyphenat}
\usepackage{wrapfig}
\usepackage{empheq}
\usepackage{textcomp}
\usepackage[caption=false]{subfig}
\usepackage{rotating}
\usepackage{wrapfig}
\usepackage{pdfpages}
\usepackage{verbatim}
\usepackage{setspace}
\usepackage{array}
\usepackage{caption}
\usepackage[pdf]{pstricks}
\usepackage{upgreek}
\usepackage{cmll}
\usepackage{latexsym}
\usepackage{braket}
\usepackage{cite}
\usepackage{epsfig}
\usepackage[left=2.4cm,top=3.3cm,right=2.4cm,bottom=3.3cm,bindingoffset=0cm]{geometry}
\usepackage[titletoc,page]{appendix}

\usepackage{soul}
\sethlcolor{yellow}

\newcommand{\beq}{\begin{eqnarray}}
\newcommand{\eeq}{\end{eqnarray}}
\newcommand{\bea}{\begin{eqnarray}}
\newcommand{\eea}{\end{eqnarray}}
\newcommand{\be}{\begin{equation}}
\newcommand{\ee}{\end{equation}}

\def\p{\varphi}

\def\de{\partial}

\def\1{\mathbbm{1}}

\def\p{\varphi}

\def\S{\mathcal{S}}

\def\Tr{\qopname\relax o{Tr}}
\def\tr{\qopname\relax o{Tr}}

\newcommand{\bx}{\mathbf{x}}
\newcommand{\btau}{\boldsymbol{\tau}}

\numberwithin{equation}{section}

\newcommand{\bulk}{\Phi}

\begin{document}

	\title{\bf{ \Large Neutron stars and phase diagram\\ in a hard-wall AdS/QCD model}}
	\author{
	 Lorenzo Bartolini$^1$\footnote{lorenzo(at)henu.edu.cn}\ , 
	 Sven Bjarke Gudnason$^1$\footnote{gudnason(at)henu.edu.cn}\ ,\\ 
	 Josef Leutgeb$^2$\footnote{josef.leutgeb(at)tuwien.ac.at}\ \  and 
	 Anton Rebhan$^2$\footnote{anton.rebhan(at)tuwien.ac.at}\\[20pt]
		{\em\normalsize
		$^1$Institute of Contemporary Mathematics, School of
  Mathematics and Statistics,}\\ {\em\normalsize Henan University, Kaifeng, Henan 475004,
  P.~R.~China}\\
		{\em\normalsize $^2$Institut
für Theoretische Physik, Technische Universität Wien,}\\ {\em\normalsize Wiedner Hauptstrasse 8-10, A-1040 Vienna, Austria
	}}
	\date{February 2022}
	\maketitle
	\thispagestyle{empty}
	\setcounter{page}{0}
	\begin{abstract}
        We study the phase diagram of (large-$N_c$) QCD using a simplistic holographic hard-wall model with a dynamical scalar field and a homogeneous Ansatz representing a smeared instanton/baryon density. The resulting phase diagram is qualitatively consistent with expectations, including a mesonic, baryonic, quarkyonic, and quark-gluon plasma phase. As in other holographic models, we also find a baryonic popcorn transition, which appears at large chemical potential as a crossover. We then evaluate the nuclear matter equation of state, which turns out to be rather stiff with a large peaked sound velocity above the conformal limit, construct corresponding neutron stars using the TOV equations, and finally use a full numerical gravity/hydrodynamics computation to extract the gravitational wave signal of neutron star mergers.
	\end{abstract}
	\newpage
	\tableofcontents
\section{Introduction}

Determining the phase diagram of QCD, the theory governing the constituents of nuclear matter at the fundamental level, is one of the most important issues in modern theoretical physics, with applications to the cosmology of the early universe, neutron stars, hadronic physics, cosmic rays, heavy-ion collision experiments and more.
The confinement of quarks and gluons into hadrons and in particular baryons is an essential feature of nature, which allows the low-energy physics to be described in terms of atoms, molecules and electrons with suitable theories at those scales.
This fact is strongly tied with asymptotic freedom of QCD, which permits to describe the high-energy processes with satisfactory precision using perturbation theory.
Unfortunately, ordinary matter is governed by the low-energy limit of QCD, where the theory runs into strong coupling and is tremendously hard to study theoretically.
Brute-force numerical computations of QCD in the version where spacetime is discretized on a lattice is so far the most reliable source of our knowledge about the phase diagram of QCD.
Unfortunately, lattice QCD works well only for small or vanishing chemical potentials, due to a technical problem known as the sign problem, which is a severe obstacle to the convergence of the numerical computations \cite{deForcrand:2009zkb,Aarts:2015tyj,Nagata:2021bru}.

In 1997 an alternative toolbox became available to theoretical physicists by the discovery of the AdS/CFT correspondence by Maldacena \cite{Maldacena:1997re}. In the original and most precise version, it relates a particular conformal field theory (CFT), namely 4-dimensional $\mathcal{N}=4$ super-Yang-Mills theory in the limit of large color number $N_c$, to supergravity in 5-dimensional anti-de Sitter (AdS) space.
The exciting property about the correspondence or duality is that the CFT at strong coupling is mapped to gravity at weak coupling, which means that perturbation theory in the bulk in a theory with one extra dimension can be used to studying the field theory living on the boundary of AdS at strong coupling.
Conformal field theories do not possess particles as we know them, but describe operators, their fall-offs, and their anomalous dimensions.
A large volume of work in the past 25 years, however, has shown that various deformations of the correspondence are possible, which has led to enormous progress in
the understanding of strongly coupled gauge theories,
see e.g.~\cite{Ammon:2015wua,Kim:2012ey} for reviews.

There are basically two ways to use the holographic correspondence to studying QCD, which are top-down or bottom-up, both of which are often referred to by now as holographic QCD.
The top-down category of holographic models are based on constructions in string theory that give rise to QCD-like theories in the low-energy limit. The most successful of those is arguably the Witten-Sakai-Sugimoto model \cite{Witten:1998zw,Sakai:2004cn}, where flavor and color branes geometrically describe the theory of strong interactions with chiral quarks resulting from strings with one end point on a flavor brane and one end point on the color branes, whereas gluons have both end points on the color branes and are thus adjoint fields. In the confined phase, the flavor branes and anti-branes corresponding to left and right-handed quarks merge in the cigar-shaped bulk geometry and thus geometrically explain nonabelian chiral symmetry breaking.

A simpler holographic setup is inspired by the same type of geometric construction and a particularly simple model considered originally by Polchinski and Strassler where
the AdS space is cut off at a finite value of the radial coordinate
\cite{Polchinski:2001tt,Boschi-Filho:2002wdj,Polchinski:2002jw,Boschi-Filho:2002xih}.
Meson and baryons were then subsequently added to the model, providing the basis of perhaps the simplest possible phenomenologically viable holographic description of QCD and hadron physics \cite{deTeramond:2005su,Erlich:2005qh,DaRold:2005mxj,Hirn:2005nr,Karch:2006pv}, which
we shall take as basis for our study.
In contrast to the Witten-Sakai-Sugimoto model, where deconfinement and chiral symmetry restoration can be set up to occur at different temperatures \cite{Aharony:2006da,Horigome:2006xu},
the simplest version of the hard-wall model has them locked together.
In order to enrich the phase structure of the theory also in the hard-wall models, we implement a so-called double hard-wall condition, where the wall for the gluons is kept at the hard-wall, but allowing for the flavor gauge and scalar fields to end on a second hard-wall, thus giving the possibility of separately adjusting the deconfinement transition. 

As is well-known, the low-energy limit of the strong interactions is sufficiently well described by only two flavors of quarks, namely the up and the down quarks -- enough for describing proton, neutrons and pions. The flavor symmetry of this low-energy sector is thus $SU(2)$, which in 4-dimensional spacetime possesses a topological degree, which counts the number of instantons. In the holographic context, the instantons in the bulk are identified as the baryons of the theory.

In this paper, we are interested in the part of the QCD phase diagram where the chemical potential is finite to large. 
For this reason, we employ a homogeneous Ansatz, describing the instantons in the approximation suitable for large densities or equivalently finite/large chemical potential. By large densities, we mean larger than the saturation density of nuclear matter but below the asymptotic, perturbative regime.
The phase diagram we find is rather rich, phenomenologically, and is consistent with common lore for the different phases and their approximate placement in the diagram \cite{McLerran:2007qj,Fukushima:2013rx}.
Indeed we find mesonic, baryonic and quarkyonic (baryonic popcorn \cite{Kaplunovsky:2012gb}) phases as we increase the chemical potential for small but finite temperatures. 
For large temperatures, the model loses the hard-wall as it moves behind the black hole horizon in the bulk geometry and the theory is thus in the chiral symmetry restoration phase.
We nevertheless find a new speculative phase transition at large temperatures for relatively large chemical potential, where a kind of quarkyonic phase appears in our model -- in the sense that deconfined quarks and baryons coexist -- reminiscent of the quarkyonic phase found in the Witten-Sakai-Sugimoto model in \cite{Kovensky:2020xif}.

We use the model in the baryonic phase to calculate the equation of state, which for a range of densities of order of a few times the saturation density for nuclear matter is a difficult region for other types of models to make predictions for.
Indeed, traditional nuclear models are trustable only at lower densities and perturbative QCD is trustable only at much larger densities than can be realized in neutron stars.
Since already simple hard-wall holographic QCD models have proven surprisingly
successful in describing low-energy hadron physics \cite{Kim:2012ey,Leutgeb:2021bpo}, it is interesting to explore what they are predicting for the equation of state in exactly such range of densities. 
The equation of state we find in our simple model is rather stiff compared to other holographic models in the literature\footnote{See also \cite{Jarvinen:2021jbd,Hoyos:2021uff} for recent reviews on holographic modeling of neutron stars.} \cite{Ishii:2019gta,Kovensky:2021kzl,BitaghsirFadafan:2019ofb}; at any rate, the behavior of the speed of sound having a large peak before asymptotically tending towards the conformal value $c^2_{\rm CFT}=1/3$ is a rather welcome feature.
Whereas perturbative quark matter that arises at sufficiently high baryonic densities has a speed of sound below the conformal value,
a stiffer equation of state is required for the density ranges of neutron stars' cores to reach the observed masses of at least $2M_{\odot}$, so that a peak structure at densities higher than saturation is what is expected \cite{McLerran:2018hbz}.

While our treatment does not take care of nuclear matter at lower energies and would need substantial refinements for including the outer regions of a neutron star, we also use the obtained equation of state to simulate a neutron star merger of two neutron stars of $1.4$ solar masses each at a separation distance of 45 kilometers and calculate the gravitational wave spectrum using a full-fledged numerical gravity and hydrodynamics code.

The paper is organized as follows.
In sec.~\ref{sec:action}, we set up the action of the model and introduce our notation and conventions, whereas in sec.~\ref{sec:hard-wall}, we adapt the model to the case of finite temperature, introduce a generalization of the double hard-wall as well as a dynamical scalar. In sec.~\ref{sec:homogeneous}, we discuss baryonic matter in the confined and deconfined phases and finally, in sec.~\ref{sec:phase}, we calculate the entire phase diagram of the model. In sec.~\ref{sec:neutron}, we consider neutron stars and their gravitational waves. Finally, we conclude the paper with a discussion and an outlook in sec.~\ref{sec:conclusions}. We have delegated details of the speed of sound at large densities, the single instanton used to calibrate the model, boundary conditions and the stress-energy tensor to the appendices \ref{app:soundspeed}, \ref{app:baryon}, \ref{app:BC} and \ref{app:stress-energy}, respectively.

\section{Action and notation}\label{sec:action}
	
	Largely following the setup and notation of \cite{Pomarol:2007kr,Domenech:2010aq},
	the background geometry is taken to be that of a slice of AdS$_5$ ending at bulk coordinate $z_0$ with curvature scale indicated by $L$. We will work in the mostly minus convention (when we do not Wick rotate) so that the metric assumes the form:
	\beq
	ds^2 = \frac{L^2}{z^2}\left(dx_\mu dx^{\mu} -dz^2\right).
	\label{eq:metric}
	\eeq
	The flavor field content is given by the presence of two $U(2)$ gauge vectors, $\mathcal{L}_M(x,z), \mathcal{R}_M(x,z)$, dual to the left and right-handed quark currents and a bi-fundamental complex scalar $\Phi(x,z)$ dual to the order parameter of chiral symmetry breaking. The action can be divided into three main contributions: a gauge part containing Yang-Mills-like terms, a Chern-Simons term to account for flavor anomalies, and a piece containing kinetic and interaction terms for the scalar. The minimal action reads \cite{Pomarol:2007kr,Domenech:2010aq}:
	\beq
	S &=& S_g+S_{CS}+ S_{\Phi},\\\label{Sgauge}
	S_{g}&=& -\frac{M_5}{2}\int d^4xdz\; a(z)\left[\Tr\left(L_{MN}L^{MN} \right)+\frac{1}{2}\widehat{L}_{MN}\widehat{L}^{MN} + \left\{R\leftrightarrow L\right\}\right],\\\label{SCS}
	S_{CS}&=& \frac{N_c}{16\pi^2}\int d^4xdz\;\frac{1}{4}\epsilon_{MNOPQ}\widehat{L}_M\left\{\Tr\left[L_{NO}L_{PQ}\right]+\frac{1}{6}\widehat{L}_{NO}\widehat{L}_{PQ} -\left\{R\leftrightarrow L\right\}\right\},\\
	S_{\Phi}&=& M_5 \int d^4xdz\; a^3(z)\left\{\Tr\left[\left(D_M\Phi\right)^\dag D^M\Phi\right] -a^2(z)M^2_{\bulk}\Tr \left[\Phi^\dag\Phi\right]\right\},
	\eeq
	where $a(z) = \frac{L}{z}$, 
	$L_M=L_M^a\frac{\tau^a}{2}$ and $\widehat{L}_M$ are the $SU(2)$ and $U(1)$ parts of the $U(2)$ field $\mathcal{L}_M$, 
	\beq
	\mathcal{L}_M = L_M^a \frac{\tau^a}{2} + \widehat{L}_M\frac{\mathds{1}}{2},
	\eeq
	whose field strength is
	$\mathcal{L}_{MN} = \de_{M}\mathcal{L}_{N} - \de_{N}\mathcal{L}_{M} -i\left[\mathcal{L}_M,\mathcal{L}_N\right]$ and analogously $\mathcal{R}_{MN}$ is the field strength for the field $\mathcal{R}_M$. $D_M\Phi$ is the covariant derivative defined by
	\beq
	D_M\Phi = \de_M \Phi -i\mathcal{L}_M\Phi +i \Phi \mathcal{R}_M.
	\eeq
	For spacetime indices we use instead the following labels: capital latin letters starting from $M,N,\ldots$ run over all four-dimensional space plus time, latin letters starting from $i,j\ldots$ run instead over the three dimensions of space transverse to the bulk direction. Finally, greek letters starting from $\mu,\nu,\ldots$  run over the three dimensions of space labeled by $i,j,\ldots$, plus time:
	\beq
	M,N,\ldots = {0,1,2,3,z},\qquad i,j,\ldots={1,2,3},\qquad \mu,\nu,\ldots ={0,1,2,3}.
	\eeq
	
	Requiring that the vector-vector correlator has the same asymptotic short-distance behavior as in QCD fixes $M_5 L=N_c/(12\pi^2)$ (in \cite{Erlich:2005qh} this parameter is denoted by $1/g_5^2).$ 
	
	To break chiral symmetry, the following boundary conditions are imposed in \cite{Pomarol:2007kr,Domenech:2010aq} on the IR brane:
	\beq
    \left(L_{z\mu} + R_{z\mu}\right)_{z=z_{IR}} &=& 0,\label{bdyIRLR}\\
	\left(L_{\mu} - R_{\mu}\right)_{z=z_{IR}} &=& 0.
	\eeq
	As we will see, the homogeneous Ansatz we wish to use cannot lead to a finite baryon density if both the above boundary conditions are employed. Changing boundary conditions can in principle introduce terms via the Chern-Simons action that are not present in 4D QCD: we discuss this aspect in appendix \ref{app:BC}, while for now we just keep Eq.~(\ref{bdyIRLR}), and will use another IR boundary condition on the $L_{i},R_{i}$ fields to fix baryon number density.

	The scalar field $\Phi$, upon imposition of the boundary conditions
	\beq
	\Phi(z=z_{UV})&=&\left(\frac{z_{UV}}{z_{IR}}\right)^{\Delta^-}M_q \frac{\mathds{1}}{2};\\
	\Phi(z=z_{IR})&=& \pm \xi \frac{\mathds{1}}{2},
	\eeq
	acquires a nontrivial vacuum as given in \cite{Domenech:2010aq}: 
	\beq\label{scalarvacconf}
	\langle\Phi\rangle = \frac{z_{IR}^{2\alpha}}{z_{IR}^{2\alpha}-z_{UV}^{2\alpha}}(\xi-M_q)\left(\frac{z}{z_{IR}}\right)^{\Delta^+} \frac{\mathds{1}}{2}+ \frac{1}{z_{IR}^{2\alpha}-z_{UV}^{2\alpha}}(z_{IR}^{2\alpha}M_q-z_{UV}^{2\alpha}\xi)\left(\frac{z}{z_{IR}}\right)^{\Delta^-}\frac{\mathds{1}}{2},
	\eeq
	that once the $z_{UV}\rightarrow0$ limit is taken reduces to
	\beq
	\langle\Phi\rangle = (\xi-M_q)\left(\frac{z}{z_{IR}}\right)^{\Delta^+} \frac{\mathds{1}}{2}+M_q\left(\frac{z}{z_{IR}}\right)^{\Delta^-}\frac{\mathds{1}}{2} = v(z)\frac{\mathds{1}}{2},
	\eeq
	with $\Delta^{\pm} =2\pm \sqrt{4 + L^2 M_{\bulk}^2}= 2\pm \alpha$.
	
	Here it was chosen to have the boundary condition as 
	\beq
	\Phi(z_{IR}) = +\xi \frac{\mathds{1}}{2},
	\eeq
	which is appropriate for the mesonic sector with both $M_q,\xi>0$ but, as shown in \cite{Pomarol:2007kr}, there also exists a sector with the minus sign: it is in this sector that the single baryon configuration exists. However, the homogeneous nuclear matter Ansatz will not need this sign change: we will also discuss this topic in Appendix \ref{app:BC}.
	
	As can be seen, in the usual scenario in which $L^2 M_{\bulk}^2=-3$ (that is, for $\alpha=1$), the vacuum of the scalar reduces to the configuration 
	\beq
	v(z) =  M_q\frac{z}{z_{IR}}+(\xi-M_q)\left(\frac{z}{z_{IR}}\right)^{3},
	\eeq
	where we see the explicit dependence on the infrared boundary condition of the parameter multiplying the cubic term (holographically dual to the chiral condensate).

	\subsection{Infrared boundary condition from a boundary action}
	
	In the previous sections we treated $\xi$ as a free parameter that is determined just by fitting some phenomenological data. As noted, in the chiral limit $M_q=0$ the parameter $\xi$ plays the role of the chiral condensate $\sigma$. If the quark mass is nonvanishing, then it can simply be regarded as the IR boundary value of the scalar field, dual to a combination of quark mass and chiral condensate.

However, following \cite{DaRold:2005vr} the parameter $\xi$ can be thought of as originating from a variational principle, by minimizing the energy with the addition of an IR localized boundary term to allow for a nonvanishing $\xi$:
\beq\label{eq:SIR}
\S_{IR} = \frac{m_b^2}{2} \xi^2 -  \lambda \xi^4.
\eeq
Let us now analyze what happens in the simpler scenario in which $M_{\bulk}^2L^2=-3$, $M_q=0$, $\alpha = 1$; with these parameters the vacuum of the scalar field is given by 
\beq
\langle \Phi\rangle = \xi \left(\frac{z}{z_{IR}}\right)^{3} \frac{\mathds{1}}{2},
\eeq
so that in the vacuum, $\xi$ is stabilized to
\beq\label{eq:xivac}
\xi_0^2=\frac{1}{4\lambda}\left(m_b^2-3M_5/L\right),
\eeq
due to the IR potential.
Also the value $\xi=0$ makes the energy stationary, but it is a local maximum.

The parameters $\lambda,m_b$ determine this value, but usually only $\xi_0$ is relevant, so it is simply fitted, and the former parameters are traded away.
However, we note that in nuclear matter the situation can and will change, as the coupling between the scalar and the gauge fields can introduce corrections to the stable value of $\xi$, potentially restoring chiral symmetry for certain combinations of the other parameters (chemical potential $\mu$ and temperature $T$): we will discuss this possibility in subsequent sections, after introducing suitable Ans\"atze to account for the presence of baryonic matter in the bulk.

	\section{Hard-wall model at finite temperature}\label{sec:hard-wall}
	\subsection{Hawking-Page transition}
\begin{figure}
		\centering	
		\includegraphics[width=0.25 \linewidth]{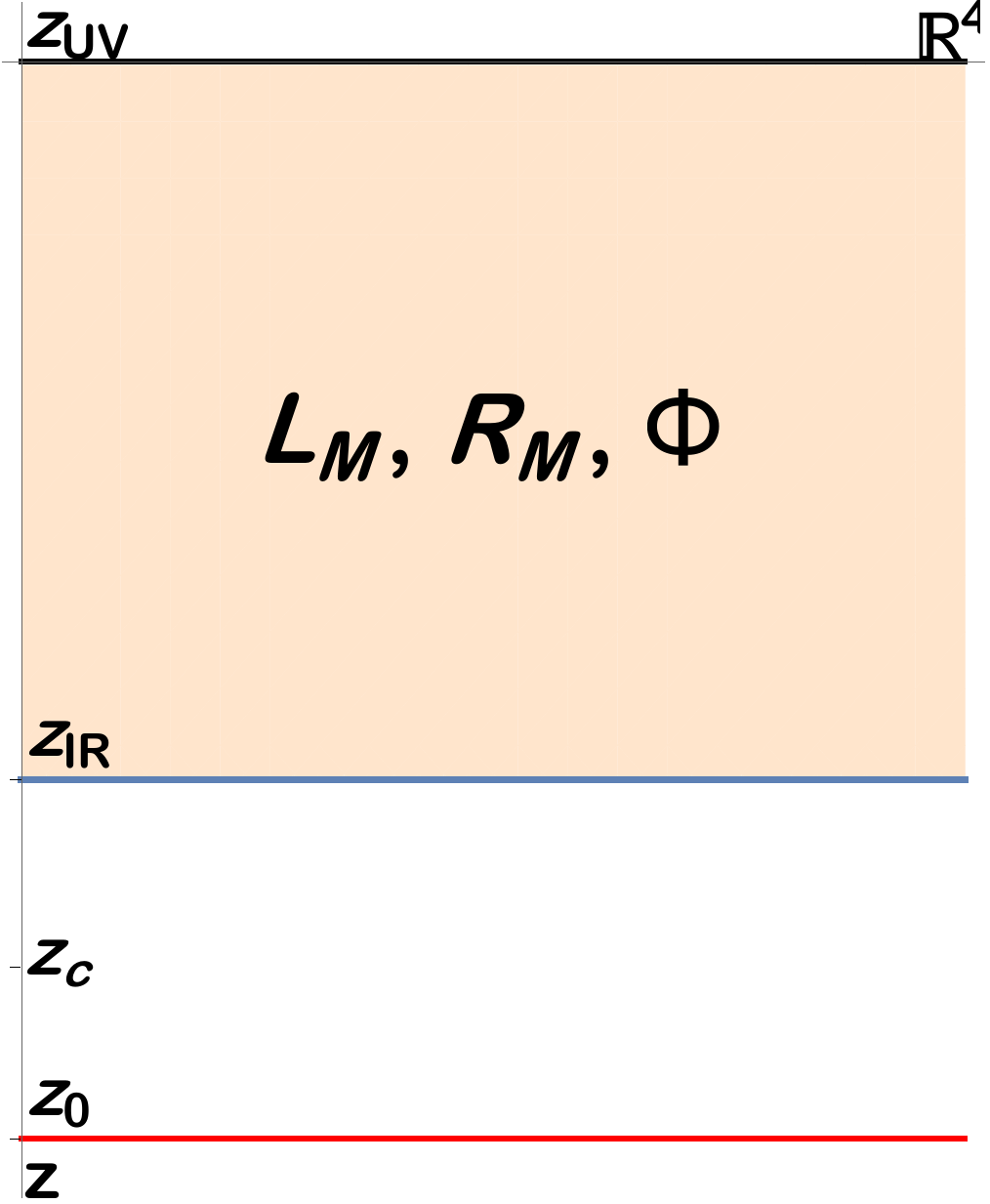}
		\includegraphics[width=0.25 \linewidth]{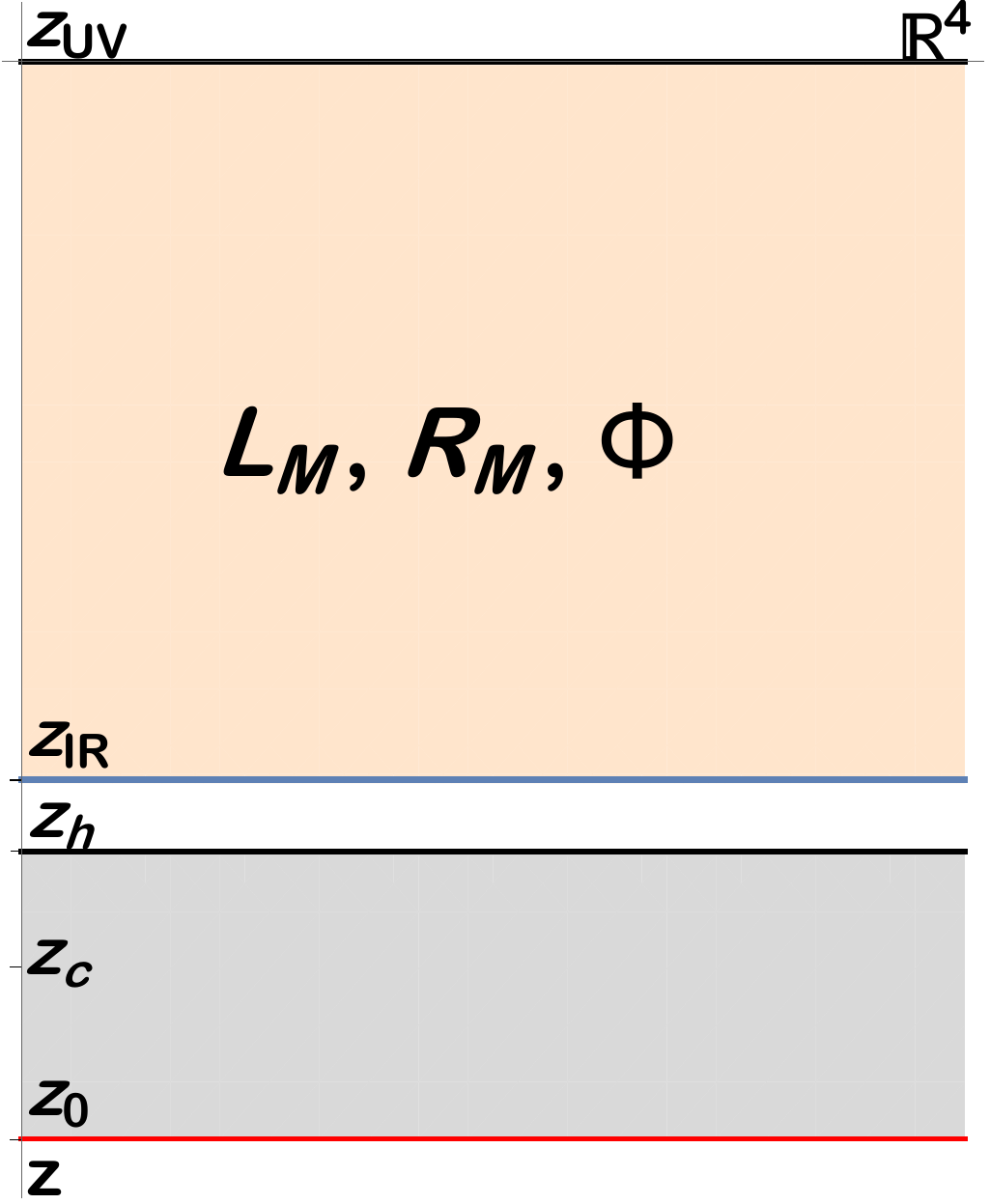}
		\caption{On the left: a sketch of our extended hard-wall model, with two separate cutoffs, $z_{IR}$ for the flavor gauge fields and $z_0$ for the bulk geometry formed by confined gluons. On the right: the model after the Hawking-Page transition. The value of $z_h$ determines the temperature, deconfinement happens when $z_h<z_c$, while chiral symmetry is restored when $z_h<z_{IR}$.}
		\label{fig:hard-wall}
	\end{figure}

	As shown in \cite{Herzog:2006ra}, the deconfinement transition happens via a Hawking-Page transition from the cutoff thermal AdS geometry to the AdS black hole geometry  described by the metric (not yet continued to Euclidean signature)
	\beq
	ds^2 = \frac{L^2}{z^2}\left(f(z)dt^2-dx_i^2-\frac{dz^2}{f(z)}  \right),\qquad f(z)= 1-\left(\frac{z}{z_h}\right)^4.
	\eeq
	The temperature of the dual theory is determined by the periodicity of the euclidean time coordinate $0 \le \tau < 1/T$, which is unconstrained in the thermal AdS case, but fixed in the black hole case by the regularity of the near-horizon solution
	\beq
	T=\frac{1}{\pi z_h}.
	\eeq
	For a critical value $z_c$ of the horizon position, and thus for a certain temperature $T_d$, the black hole phase becomes energetically favored, hence the deconfinement transition.
	This transition depends only on the cutoff scale $z_{0}$:
	\beq
	z_c^4 = \frac{1}{2} z_{0}^4 \qquad \Rightarrow \qquad T_d = \frac{2^{1/4}}{\pi z_{0}}.
	\eeq
	Here $z_0$ is the cutoff for the geometry: we labeled it $z_{IR}$ in the previous section, here we introduce this new notation in order to not make confusion in the next section, so for the moment we have two names for two quantities, $z_{IR}$ (the cutoff for the flavor fields) and $z_0$ (the cutoff for the geometry), that are usually taken to coincide, see Fig.~\ref{fig:hard-wall}.
	
	This evaluation also neglects the role of the vacuum configuration of the scalar $\Phi$: its vacuum energy changes the critical horizon coordinate to favor the confined phase and introduces a dependence on $z_{IR}$ in the expression for $z_c$: however the trace of the scalar is proportional to $N_f$ so it would be of the same order as the backreaction of the fields on the geometry, a correction we are neglecting in the present work.

	\subsection{Double hard-wall model}
	
	One issue with the usual hard wall approach is that, as can be seen by the previous section on deconfinement, the critical value $z_c$ of the black hole horizon at which the phase transition happens is always lower than the cutoff $z_0=z_{IR}$.
	The horizon of the black hole, as soon as it appears, hides the IR brane on which the spontaneous breaking of chiral symmetry is realized by the boundary conditions of the gauge and scalar fields: this implies that as soon as the theory undergoes a deconfining transition, it also unavoidably restores chiral symmetry. 
	However, in general these phase transitions can be widely separated \cite{Evans:2020ztq}.
	To have a richer phase diagram with separate confined, deconfined, and chiral symmetry restored phases, we postulate that the geometrical cutoff and the gauge cutoff do not coincide, so we keep the notation $z_{IR}$ for the cutoff of the flavor gauge fields propagation which we permit to be smaller than $z_0$ where the geometry ends.
	
	This kind of construction looks artificial, but an analogous configuration is actually realized in the top-down holographic QCD model of Witten-Sakai-Sugimoto when the flavor D-branes are not chosen to be antipodal: in this configuration, the flavor gauge fields that live on the branes also do not exist for all values of the radial coordinate.
Since the Hawking-Page transition is sensitive only to $z_0$, to have 
	$T_d\not=T_c$
	we need to have 
	\beq
	z_{0} > 2^{1/4}z_{IR}.
	\eeq
	Since our critical temperature corresponds to a critical $z_c = 2^{-1/4} z_0$, having a higher value of $z_0$ than in \cite{Herzog:2006ra} (since their $z_0$ corresponds now to our $z_{IR}$) means ending up with a lower temperature for the Hawking-Page transition
	\beq
	T_d = \frac{2^{1/4}}{\pi z_0}<\frac{1}{\pi z_{IR}}\approx 102.6 \text{ MeV},
	\eeq 
	departing even more from a phenomenologically accurate value, if $z_{IR}$ is fixed by
	the mass of the rho meson to be 
	\beq
	z_{IR}=L \qquad ; \qquad L^{-1}=323\mathrm{MeV},
	\eeq
     as is usually done.
	An alternative way to look at this choice without invoking a ``double wall'' structure, is simply to take the deconfinement temperature to be a free parameter, disregarding the dynamics of the deconfining Hawking-Page transition and instead imposing the change in geometry by hand: we will take this approach in drawing the phase diagram, plotting temperatures as ratios of the critical temperature $T_c=(\pi z_{IR})^{-1}$.
	
	This choice will not affect the main results of this paper, namely the equation of
	state of cold nuclear matter and the properties of neutron stars, whose typical temperatures are far below the scale of $T_c$ and $T_d$, for which we shall consider the model at $T=0$.
	The benefit of having a setup where the deconfinement temperature $T_d$ is different from the critical temperature $T_c$ rather lies in reproducing the analogous freedom that is found in the Witten-Sakai-Sugimoto model with non-antipodal D-branes, thereby permitting to see whether any
	of the features of the latter can be reproduced in our much simpler model.
	
\subsection{Scalar thermal vacuum}
	We have already shown in Eq.~(\ref{scalarvacconf}) the vacuum configuration of the scalar for nonvanishing $M_q$ and $\xi$ in the truncated AdS$_5$ geometry. Since we want to study the theory also in the deconfined phase, we need to find the vacuum configuration also for the thermal black hole geometry, for the case in which the black hole horizon lies beyond the IR-brane.
	
	Parametrizing again $\langle\Phi\rangle = v(z)\mathds{1}$, the equation of motion for the scalar alone is modified by the presence of blackening factors $f(z)$ as
	\beq
	\de_z\left(f(z)a^3(z)\de_z v(z)\right) - M_{\bulk}^2 a^5(z)v(z)=0\qquad;\qquad f(z)=1-\frac{z^4}{z_h^4},
	\eeq
	keeping the same boundary conditions provided in \cite{Domenech:2010aq}.
	The presence of the blackening factors modifies the configuration (\ref{scalarvacconf}) by the introduction of hypergeometric functions, so that a general solution 
	reads
	\beq
v(z)&=&C_1(-1)^{\frac{2 - \alpha}{4}}  \left(\frac{z}{z_h}\right)^{2 - \alpha}
	 {}_2F_1\left[\frac{1}{2} - \frac{\alpha}{4},\frac{1}{2} - \frac{\alpha}{4}, 
	1 - \frac{\alpha}{2}, \left(\frac{z}{z_h}\right)^4\right]\nonumber\\
	 &+&C_2(-1)^{\frac{2 + \alpha}{4}} \left(\frac{z}{z_h}\right)^{2 + \alpha}
	 {}_2F_1\left[\frac{1}{2} + \frac{\alpha}{4},\frac{1}{2} + \frac{\alpha}{4}, 
	1 + \frac{\alpha}{2}, \left(\frac{z}{z_h}\right)^4\right],
	\eeq
with $C_1,C_2$ integration constants (but potentially dependent on $T$ through $z_h$) to be fixed by the boundary conditions. Here we provide the full configuration for our boundary conditions: we employ the following shorthand notation
\beq
F_\pm(z) \equiv \quad_2F_1\left(\frac{2 \pm \alpha}{4},\frac{2 \pm \alpha}{4},\frac{2 \pm \alpha}{2},\frac{z^4}{z_h^4}\right).
\eeq
Then the vacuum of the scalar in the deconfined background is given by:
\beq
v(z,T) &=&C^{-1}\left(\xi F_-(z_{UV}) -M_q  F_-(z_{IR})\right)   \left(\frac{z}{z_{IR}}\right)^{2+\alpha}    z_{IR}^{2\alpha}F_+(z) \nonumber\\& +&C^{-1}\left( M_q z_{IR}^{2\alpha} F_+(z_{IR})  -\xi z_{UV}^{2\alpha} F_+(z_{UV})\right)\left(\frac{z}{z_{IR}}\right)^{2-\alpha} F_-(z)  ,\label{vacuumthermal}
\eeq
with the constant $C$ being
\beq
C\equiv z_{IR}^{2\alpha}F_-(z_{UV})F_+(z_{IR})-z_{UV}^{2\alpha}F_-(z_{IR})F_+(z_{UV}).
\eeq

\section{Homogeneous baryonic matter}\label{sec:homogeneous}
From the holographic point of view baryons are solitonic configurations in the flavor gauge fields: to build a many-instantons configuration accounting for interactions and minimization of the energy with respect to the moduli of the solitons is extremely challenging, so usually an approximation scheme is necessary to extract some qualitative result from the model. 

Another possibility is to take the fields to only depend on $z$ to begin with, instead of starting with individual baryons and then performing some averaging.
A configuration like this\footnote{The same Ansatz has been employed to describe dense nuclear matter within the top-down model of Witten-Sakai-Sugimoto, see \cite{Li:2015uea}} would, of course, be far from reality when densities are low, but as the density increases the system tends to look homogeneous up to very short distances in $\mathds{R}^3$.

The benefit of using this approximation is threefold: first, it reduces the number of variables in our equations of motion, secondly it decouples the nonabelian part of the scalar field $\Phi$ from the baryonic matter, hence allowing us to only turn on its $U(1)$ term, and lastly it allows us to easily solve the equations of motion numerically without fixing the configuration of certain fields, so that the resulting field configuration will be a true classical solution of the theory.
	
	The gauge fields and the scalar are taken to depend only on $z$, so as to reproduce a distribution of a large number of baryons smeared in flat three-dimensional space. This requirement alone, together with the condition (\ref{bdyIRLR}) for the gauge fields, and with $M_q = m\mathds{1}$ for the scalar, enforces the following form \cite{Domenech:2010aq}:
	\begin{align}
	L_z = R_z &=0,\\
	\label{hansatz}
	L_i = -R_i = -H(z)\frac{\tau^i}{2} \qquad &; \qquad \widehat{L}_0 = \widehat{R}_0 = \widehat{a}_0(z),\\
	\Phi = \omega_0&(z)\frac{\mathds{1}}{2}.
	\end{align}
	The imposition of $R_i = -L_i$ is also necessary in order to obtain a nonvanishing baryon number. 
	The field strengths are then given by:
	\beq
	L_{iz} &=& -R_{iz} = H'(z)\frac{\tau^i}{2},\\
	L_{ij}&=& R_{ij} = \frac{H^2(z)}{2} \epsilon^{ijk}\tau^k, \\
	\widehat{L}_{0z} &=& \widehat{R}_{0z} = -\widehat{a}'_0(z).
	\eeq
	
	From now on we shall simplify our equations by setting 
	\beq
	z_{UV}=0, \quad z_{IR} = L = 1.
	\eeq
($z_{IR}=1$ corresponds to measuring all dimensionful physical quantities in some units essentially fixed by the rho mass, to be fitted later, whereas $L=z_{IR}=1$ can be achieved by appropriate rescalings.)

	Given the Ansatz in the previous paragraph, we can extract a one-dimensional effective Lagrangian density from the model. It is given by
	\beq
	\mathcal{L} &=& \mathcal{L}_g + \mathcal{L}_{CS} + \mathcal{L}_{\Phi},\\
\mathcal{L}_g 	&=&- M_5 a(z) \left[3H^4 + 3H^{'2} - \widehat{a}_0^{'2}\right],\\
	\mathcal{L}_{CS}&=&\frac{3N_c}{8\pi^2}\widehat{a}_0 H^2H',\\
	\mathcal{L}_\Phi &=& -M_5a^3(z)\left[\frac{3}{2}H^2\omega_0^2 +\frac{1}{2}M_5\omega'^2_0 +\frac{1}{2} M_{\bulk}a^2(z) \omega_0^2\right].
	\eeq
	From this Lagrangian density we can obtain the grand-canonical potential via the holographic correspondence as
	\beq
	&&\Omega = -V \int dz\; \mathcal{L}.
	\eeq
	
	\paragraph{Equations of motion}
	The set of equations of motion for the $z-$dependent fields of the homogeneous Ansatz are easily obtained as:	
	\begin{align}
	&6M_5 \de_z\left(a(z)H'(z)\right) -12a(z) M_5H^3(z) -\frac{3N_c}{8\pi^2}\widehat{a}'_0(z)H^2(z) - 3M_5a^3(z)H(z)\omega_0^2=0,\label{eqH}\\
	&-2M_5\de_z\left(a(z)\widehat{a}'_0(z)\right) +\frac{3N_c}{8\pi^2}H^2(z)H'=0,\label{eqa0}\\
	&M_5\de_z\left(a^3(z)\omega'_0\right) - 3M_5a^3(z)H^2(z)\omega_0-M_5M_{\bulk}^2a^5(z)\omega_0=0.\label{eqomega}
	\end{align}
 	The boundary conditions in the UV descend from 
 	\beq
 	L_{\mu}|_{UV} &=& R_{\mu}|_{UV} = 0;\\
 	\Phi|_{UV} &=& \left(\frac{z_{UV}}{z_{IR}}\right)^{\Delta^-} M_q \frac{\mathds{1}}{2},
 	\eeq
 	and are thus translated into Dirichlet conditions for the fields of the homogeneous Ansatz:
 	\beq
 	H(z_{UV}=0) &=& 0,\label{eq:HUV}\\
 	\omega_0(z_{UV}=0) &=& 0.
 	\eeq
	We also need to specify conditions for the IR boundary: for the fields $\Phi(z)$ and $\widehat{a}_0(z)$ we still make use of the original boundary conditions, that is
	\beq
	\Phi(z_{IR}=1) &=& \xi\frac{\mathds{1}}{2},\\
	\widehat{L}_{z\mu}(z_{IR}=1) &=& \widehat{R}_{z\mu}(z_{IR}=1) = 0.
	\eeq
	As mentioned earlier, the only IR condition we have imposed for the field $H(z)$ is automatically satisfied by construction. We still need one more boundary condition for the field $H(z)$ to be able to integrate the differential equation: this freedom reflects the freedom that we have to choose an arbitrary baryon number for the constructed Ansatz. The baryon number is obtained as the topological charge
	\beq
	B &=& \frac{1}{32\pi^2} \int d^3x dz\; \epsilon_{MNPQ} \tr \left[L^{MN}L^{PQ}-R^{MN}R^{PQ}\right] \nonumber\\
	&=& \frac{3V}{4\pi^2} \int^{1}_0 dz\; H'H^2\nonumber\\
	&=& \frac{V}{4\pi^2}\left[H^3\right]_0^1.
	\eeq
	Since $H(0)=0$, the value of $B$ can be fixed by $H(1)$: since we are working with an infinite volume it is more meaningful to trade the global topological charge $B$ with its density $d$, obtaining the following boundary conditions for $H(1)$ at a given value of $d$, and for the other fields
	\beq
	H(z_{IR}=1) = \left(4\pi^2 d\right)^{\frac{1}{3}}\label{eq:HIR}\quad;\quad
	\widehat{a}'_0(z_{IR}=1) = 0\quad;\quad
	\omega_0 = \xi.
	\eeq
	Note that the field $\widehat{a}_0$ appears only through its derivative: we will thus be able to determine its shape up to an additive constant. This is not unexpected, and is in agreement with the holographic interpretation of such constant, that is, the chemical potential $\mu_B$: it can easily be understood by inspection of the Chern-Simons action, which reduces to a coupling between $\mu_B$ and the baryon number (density). We are thus free to choose this value, as we are free to choose the chemical potential we are working at.
	Furthermore, since Eq.~(\ref{eqa0}) contains the baryon number density (which reduces to a boundary term upon integration), it can be immediately integrated and solved for $\widehat{a}'_0(z)$: 
	\beq
	\widehat{a}'_0(z)= \frac{N_c}{16\pi^2M_5}\frac{H^3(z)}{a(z)} + \frac{C}{a(z)}. \label{eqdea0}
	\eeq
	The integration constant $C$ can be fixed using the Neumann boundary condition at $z=z_{IR}=1$ for $\widehat{a}_0(z)$, and then the Dirichlet condition for $H(z)$:
	\beq
	C = -\frac{N_c}{16\pi^2 M_5} H^3(1) = -\frac{N_c d}{4M_5}.
	\eeq
	So after all we can solve the two equations (\ref{eqH}), (\ref{eqomega}) (using (\ref{eqdea0})) to find $H(z),\omega(z)$ and then plugging $H(z)$ into Eq.~(\ref{eqdea0}) together with the boundary condition $\widehat{a}_0(0)=\mu$ to find $\widehat{a}_0(z)$. Note however that with this notation, the integration constant $\mu$ doesn't give the physical chemical potential for baryons $\mu_B$, but is proportional to it via
	\beq
	\mu_B= \frac{N_c}{2}\mu,
	\eeq
	which will be important to take into account when fitting the free parameters to obtain phenomenologically meaningful quantities.

	\subsection{Confined phase}
	To move to the finite temperature theory we perform a Wick rotation as $t=-i\tau$: because of this, the time components of vector fields will be defined as $L_0 = iL_\tau$. Keeping the same homogeneous Ansatz as that of Eq.~(\ref{hansatz}), but trading $L_0,R_0$ for $L_\tau, R_\tau = -iL_0 = -i\widehat{a}_0 \equiv \widehat{a}_\tau$, the one-dimensional Lagrangians read
	\beq
	\mathcal{L} &=& \mathcal{L}_g + \mathcal{L}_{CS} + \mathcal{L}_{\Phi},\\
\mathcal{L}_g 	&=& i M_5 a(z) \left[3H^4 + 3H^{'2} + \widehat{a}_\tau^{'2}\right],\\
	\mathcal{L}_{CS}&=&\frac{3N_c}{8\pi^2}\widehat{a}_\tau H^2H',\\
	\mathcal{L}_\Phi &=& iM_5a^3(z)\left[\frac{3}{2}H^2\omega_0^2 +\frac{1}{2}\omega'^2_0 +\frac{1}{2} M_{\bulk}a^2(z) \omega_0^2\right].
\eeq
	Consistency of the Chern-Simons term with the other ones requires the field $\widehat{a}_\tau$ to be imaginary, so we rewrite it as $\widehat{a}_\tau= -i\widehat{a}_0$ with $\widehat{a}_0(z)$ a real field. We then define the euclidean Lagrangian $\mathcal{L}^E$ as $\mathcal{L}^E = -i\mathcal{L}$. With these prescriptions we end up with the very same form of the Lagrangian as in the zero temperature scenario, and thus with the same equations of motion: no temperature induced corrections to the baryonic matter appear in this particular phase.
	By substituting the numerical solution of the equations of motion for this phase into the action, we can compute the grand canonical potential. 
	
	\subsection{Deconfined phase}
	
	When the geometry undergoes a transition to the black hole background, the effective one-dimensional Lagrangian (and correspondingly the equations of motion) change, including additional blackening factors in some terms:
	
	\beq
	\mathcal{L}^E &=& \mathcal{L}^E_g + \mathcal{L}^E_{CS} + \mathcal{L}^E_{\Phi},\\
	\mathcal{L}^E_g 	&=&  M_5 a(z) \left[3H^4 + 3 f(z) H^{'2} - \widehat{a}_0^{'2}\right],\\
	\mathcal{L}^E_{CS}&=&-\frac{3N_c}{8\pi^2}\widehat{a}_0 H^2H',\\
	\mathcal{L}^E_\Phi &=& M_5a^3(z)\left[\frac{3}{2}H^2\omega_0^2 +\frac{1}{2}f(z) \omega'^2_0 +\frac{1}{2} M_{\bulk}^2 a^2(z) \omega_0^2\right],
	\eeq
	leading to the following set of equations:
	\begin{align}
	&6M_5 \de_z\left(f(z) a(z)H'(z)\right) -12a(z) M_5H^3(z) -\frac{3N_c}{8\pi^2}\widehat{a}'_0(z)H^2(z) - 3M_5a^3(z)H(z)\omega_0^2=0,\label{eqTH}\\
	&-2M_5\de_z\left(a(z)\widehat{a}'_0(z)\right) +\frac{3N_c}{8\pi^2}H^2(z)H'=0,\label{eqTa0}\\
	&M_5\de_z\left(f(z)a^3(z)\omega'_0\right) - 3M_5a^3(z)H^2(z)\omega_0-M_5M_{\bulk}^2a^5(z)\omega_0=0.\label{eqTomega}
	\end{align}
	Boundary conditions can be imposed as in the $T=0$ scenario. This time $z_h$ (hence temperature) enters directly into the equations of motion, so the corresponding nuclear matter solution will have temperature induced corrections: this translates in the phase diagram $T$ vs $\mu$ into the phase transition line between mesonic phase and nuclear matter phase no longer being a straight vertical line (a feature already seen for pointlike instantons).
	Numerical evaluation confirms this and also establishes the transition line remaining of first order. 

\section{Phase diagram}\label{sec:phase}

\subsection{Chiral symmetry restoration at zero temperature}

The presence of couplings between the scalar field and $H(z)$ implies that a finite density can indeed introduce corrections to the stable value of $\xi$ which is determined in the vacuum by
the IR boundary action (\ref{eq:SIR}) according to Eq.~(\ref{eq:xivac}). 

To show how this happens, let us start from a simplified picture, in which the value of $\xi$ does not affect the shape of $H(z)$: if this was the case, then the only relevant additional term in the stabilization of $\xi$ would be given by the one that explicitly contains $\omega_0(z)$
\beq
\mathcal{L}_{\Phi}^{H\omega} &=& -\frac{3}{2}M_5 a^3(z) H^2\omega_0^2\nonumber\\
&=&-\frac{3}{2}M_5\xi^2 z^3 H^2.
\eeq
Upon integration over $z$ and exploiting the assumption that $\de_{\xi}H=0$, we find the following stable value for $\xi$ that minimizes the energy:
\beq\label{stablexid}
\xi^2= \frac{1}{4\lambda}\left[m_b^2 - 3M_5\left(1+\mathcal{J}\right)\right]\qquad,\qquad \mathcal{J}\equiv \int dz z^3 H^2,
\eeq
where $\lambda$ is the self coupling of $\xi$, see eq.~\eqref{eq:SIR}.
The integral $\mathcal{J}$ depends on the density and behaves roughly as $d^{2/3}$, but in general it is sufficient to note that it is a monotonic increasing function of the baryon density. We note that in nuclear matter we then have indeed a dynamical chiral condensate. More interestingly, the additional term can in principle allow for a restoration of chiral symmetry: if the right-hand side of Eq.~(\ref{stablexid}) becomes negative, this signals that only the solution $\xi=0$ survives, and it becomes a local minimum. This would establish a critical density at which chiral symmetry is restored even at zero temperature.

\begin{figure}
    \centering
    \includegraphics[width=0.65\linewidth]{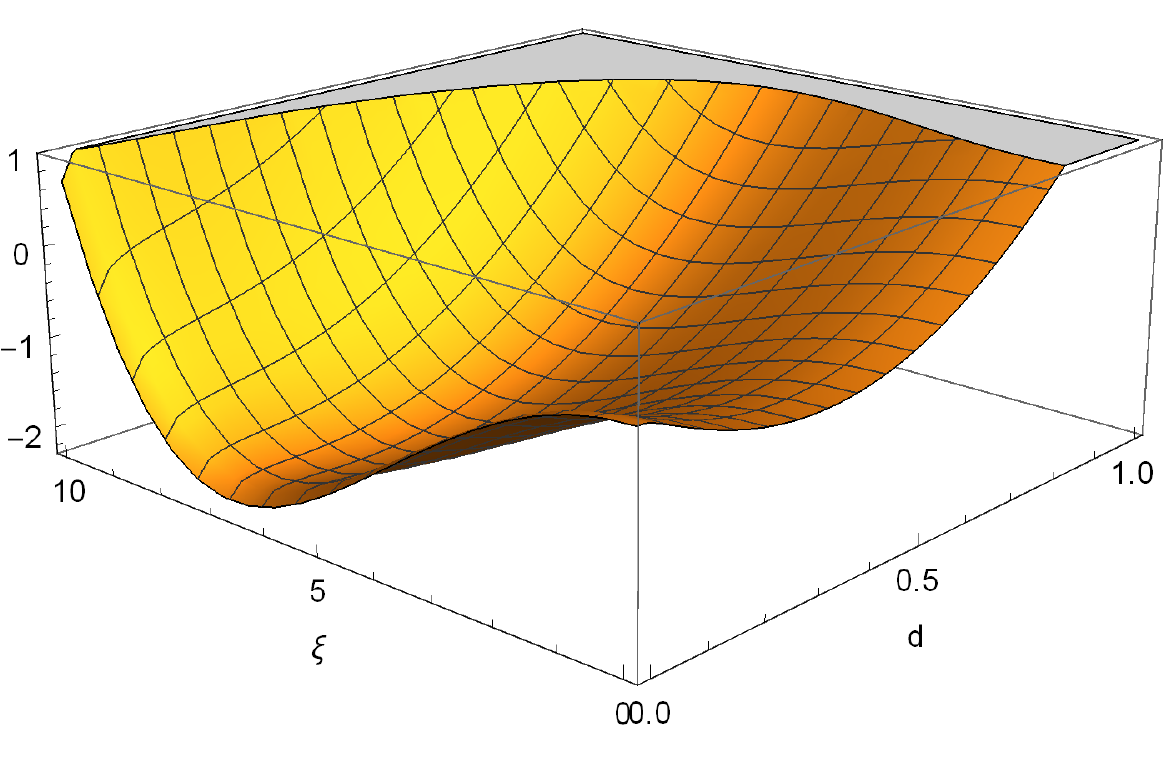}
    \caption{ A 3D plot of the grand canonical potential at $T=0$ as a function of $\xi$ and $d$, at a fixed value of $\mu=5$ for $\lambda=0.001$ and $m_b=0.512$, in units where $L=1$ (these values determine an equilibrium value $\xi=4\sqrt{2}$ and are used in one of our choices of calibration). Two local minima develop along the axes $d=0$, $\xi=0$, and $\mu$ will determine their position along the axes and which one is the global minimum. This general structure is not modified by the choice of parameters or a finite temperature, but the local minima may vary for different values of $m_b,\lambda, T$.}
    \label{fig:0Tenergy}
\end{figure} 

In reality, the function $H(z)$ also depends weakly on $\xi$ and so does $\mathcal{J}$, so the behavior of the energy as a function of $d,\xi$ has to be evaluated numerically. 

To perform the numerical analysis, we need to find a scheme to fit the free parameters of the model. At this stage, the free parameters are $L, m_b, \lambda$. The parameter $m_b$ can be traded for the value of $\xi$ since it is obtained by minimizing the energy. The most common fit for $L$ has it chosen to be $L^{-1}= 320$MeV, to reproduce the $\rho-$meson mass, while the parameter $\xi$ in the vacuum at zero temperature should be set as\footnote{Note that with respect to \cite{DaRold:2005vr} we have a definition of $\xi$ that differs by a factor of $\sqrt{2}$: this originates from the choice of the boundary conditions of the scalar field, and on the different normalization of the group structure of the fields, since we use $\tr{T^aT^b}=\frac{1}{2}\delta^{ab}$, while they use $\tr{T^aT^b}=\delta^{ab}$.} $\xi=4\sqrt{2}$: however in holographic models the fits of the mesonic sector are often at odds with observables in the baryonic sector. Since in this work we are mainly concerned with the physics of baryons, we choose a fit that reproduces baryonic observables relevant to the construction of the phase diagram and neutron stars, keeping only $\lambda$ as a free parameter. We choose our parameters to reproduce the mass of the baryon and the onset of the nuclear matter phase, independently of the chosen $\lambda$ (neglecting binding energy):
\beq\label{fit}
M_B = \frac{N_c}{2}\mu_{on} = 939.56\; \text{MeV}.
\eeq
These two values are in principle not independent as they must coincide, however since we employ the homogeneous Ansatz, the system loses track of the mass of the single baryon. 
In order to determine the mass of a single baryon, we construct a single instanton holographically in the theory, see App.~\ref{app:baryon}.

If we plot the free energy (see Fig.~\ref{fig:0Tenergy} for its complete behavior for a particular choice of $\mu$) as a function of $d,\xi$, we immediately note that the only two local minima lie on the two axes $d=0,\xi=0$, so that the two phases of the theory are one in which there are no baryons and the chiral symmetry is spontaneously broken ($d=0$ and finite $\xi$), and one in which there are baryons and the chiral symmetry is restored (finite $d$ and $\xi=0$). Which one of the two local minima is the global one depends on the chemical potential $\mu$.

To study the phases of this system we numerically find those local minima (checking that this structure of local minima holds true for each $\mu$), compare them, and determine which one is the global minimum of the function at each given value of $\mu$ (note that the value of the minimum on the $d=0$ axis is independent on $\mu$, but the one on the $\xi=0$ axis is not). 
For the purpose of this work we have chosen a nominal value of $\lambda=0.002$, a value close to the small $\lambda$ limit which gives the best results for the mass-radius relation of neutron stars among this class of calibrations.

\subsection{Chiral symmetry restoration at finite temperature}
We introduce temperature dependence by employing the deconfining geometry of the AdS black hole. Naively chiral symmetry is automatically restored as soon as the horizon of the black hole reaches $z_h=z_{IR}=1$ and hides the boundary conditions for the scalar field $\Phi(z_{IR})=\xi \mathds{1}$. 
However, before this happens, the true minimum of the energy can in principle be realized for $\xi=0$ even at lower temperatures, depending on the $\mu$ and $T$ dependence of the energy density. 
Because of the black hole metric, the IR boundary term of the action is modified as
\beq
\S_{IR}(T) =\sqrt{f(z_{IR}=1)} \left(\frac{m_b^2}{2} \xi^2 -  \lambda \xi^4\right).
\eeq
We find that this is indeed the case, and that the chiral restoration transition once again coincides with a baryonic onset, but this time the critical chemical potential $\mu_{on}(T)$ is a strictly decreasing function of the temperature. Interestingly, for $T=T_c$, the curve ends at a finite value of $\mu_{on}(T_c)$: this way our phase diagram shows a triple point as expected for large-$N_c$ QCD.

However, depending on the choice of the parameters, 
the equilibrium value of $\xi$ reaches zero for temperatures lower than $T_c$ in a second order phase transition: with our choices the critical temperatures for this dynamical chiral restoration is $T=0.939 T_c$, for the parameter set that fits the baryonic onset and baryon mass, and $T= 0.9983 T_c$ (barely visible in Fig.~\ref{fig:Phasediag1}) for the one that gives more realistic neutron stars.
This is clearly an artifact of the model, but may be interpreted as a trend towards
a weakening of the first-order phase transition around and below the value of $\mu$ at the triple point, which in real QCD should actually be a critical endpoint.

\subsection{Quarkyonic phase}
\begin{figure}[!ht]
    \centering
    \includegraphics[width=0.60\linewidth]{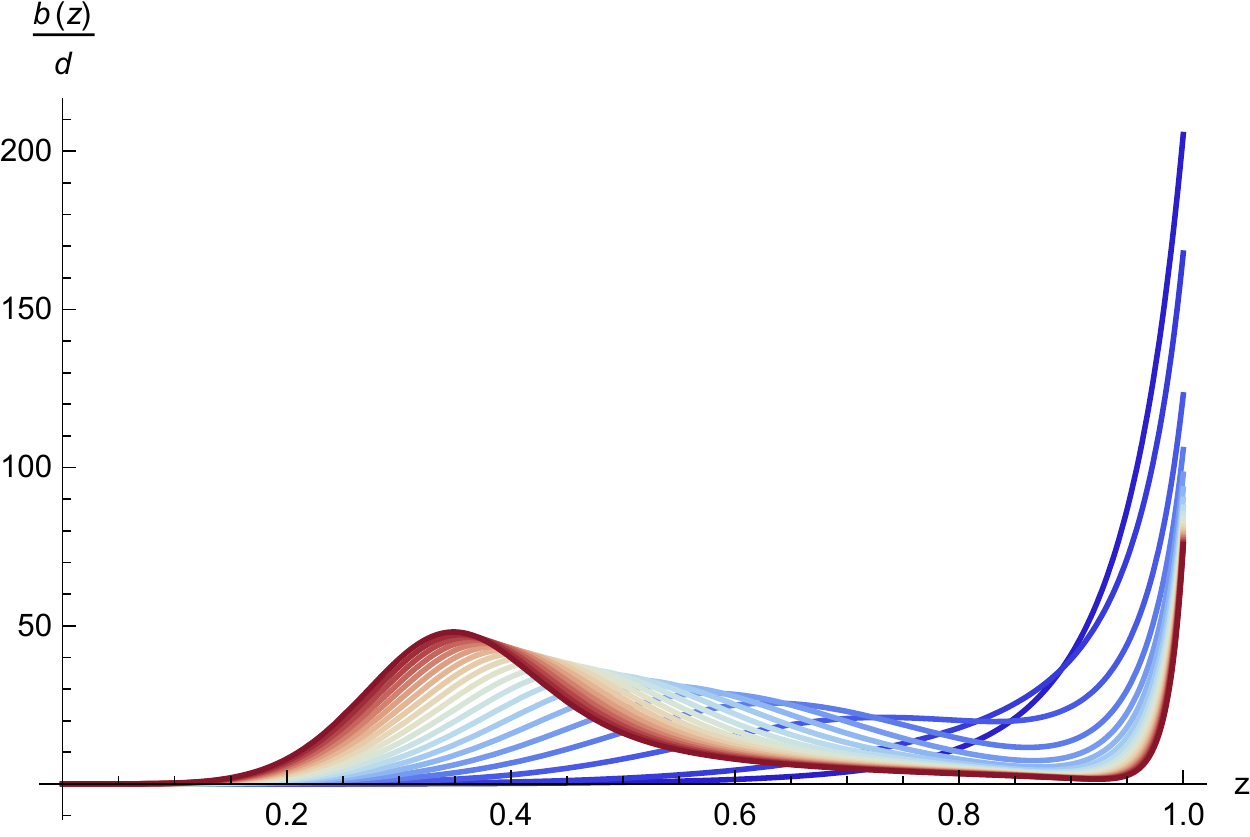}
    \caption{The baryon number density distribution in the holographic direction per unit baryon number at various densities ranging from $d=0.1$ to $d=10$ (in units where $L=1$). Colors from blue to red indicate an increase in baryon density, and the continuous development of a peak in the distribution away from the infrared wall can be clearly seen, signaling a crossover to a quarkyonic phase.}
    \label{fig:quarkyonic}
\end{figure}
We now turn our attention to the high $\mu$ region: the equilibrium density $d$ is a monotonically increasing function of $\mu$, so this means also analyzing the solutions of the equations of motion at high density. As we increase the density, we observe a continuous deformation of the baryon number density distribution, as it changes from being peaked on the infrared brane, where its boundary condition encodes the value of $d$, to a configuration where it develops a second peak at a distance from the hard-wall which is monotonically increasing with $d$ (see Fig.~\ref{fig:quarkyonic} for a set of density profiles where the peak development is manifest).
This appearance of a peak in the distribution at a finite distance from the hard wall (in general, from the position towards which the baryons are pulled by the gravitational background) is reminiscent of the ``popcorn transition'' in the non-homogeneous scenario. It was argued that the distribution in the holographic direction should be related to a spectrum of energies for the condensed baryons \cite{Rozali:2007rx}, so it is dual to a Fermi sphere. However the baryon at this level is still a classical object, neither fermion nor boson, so the fermionic nature has to come from quarks: this transition then indicates the onset of a quarkyonic phase for cold and dense nuclear matter \cite{Kaplunovsky:2012gb}. The nuclear homogeneous matter in our model exhibits the same feature, performing a (continuous) ``popcorn transition'',
where the distribution of baryon charge density is gradually pushed away from the infrared brane.
  This continuous distribution is in contrast to the case of individual layers of baryons considered in \cite{Kaplunovsky:2012gb} where the position modulus in the holographic direction is determined dynamically.

Because this is not a transition that we determine by comparing order parameters, it should be regarded as a crossover\footnote{For a different holographic realization of the quarkyonic phase, see also \cite{Chen:2019rez}.}, with the exact value of $\mu$ at which it is happening being set by our choice of the identification of when we consider this ``popcorn transition'' to have happened. Note that this would not be the case if we were working within the single-instanton approximation, since in that case we would be able to determine whether a single layer or a double layer of instantons is favored at a given density.
The choice we employed in this work to draw a crossover line in the phase diagram is to declare the transition to happen when the baryon density distribution is no longer a monotonic function of $z$, that is, when it develops a local maximum far from the infrared wall.
However, the presence of this phase has no impact on the physics of neutron stars in our model, which
we study below, since the high densities at which this crossover happens are never reached in their cores.

In \cite{Kaplunovsky:2012gb} the baryonic popcorn transition was proposed as a true
phase transition from nuclear matter to quarkyonic matter, whereas in our case these phases
are strictly speaking one. In fact, a continuity between the two has been suggested
by the lattice study of \cite{Philipsen:2019qqm}. A characteristic property of quarkyonic
matter has been argued in \cite{McLerran:2018hbz} to be a steep rise of the speed of sound
after baryon on-set to values of order one, above the conformal value $c_s^2=1/3$. As we shall
show next, this indeed happens in our model quickly after baryon on-set, already before our analog of the popcorn transition, with implications for neutron stars like in \cite{McLerran:2018hbz}.

\subsection{Speed of sound}
\begin{figure}[!ht]
    \centering
    \includegraphics[width=0.49\linewidth]{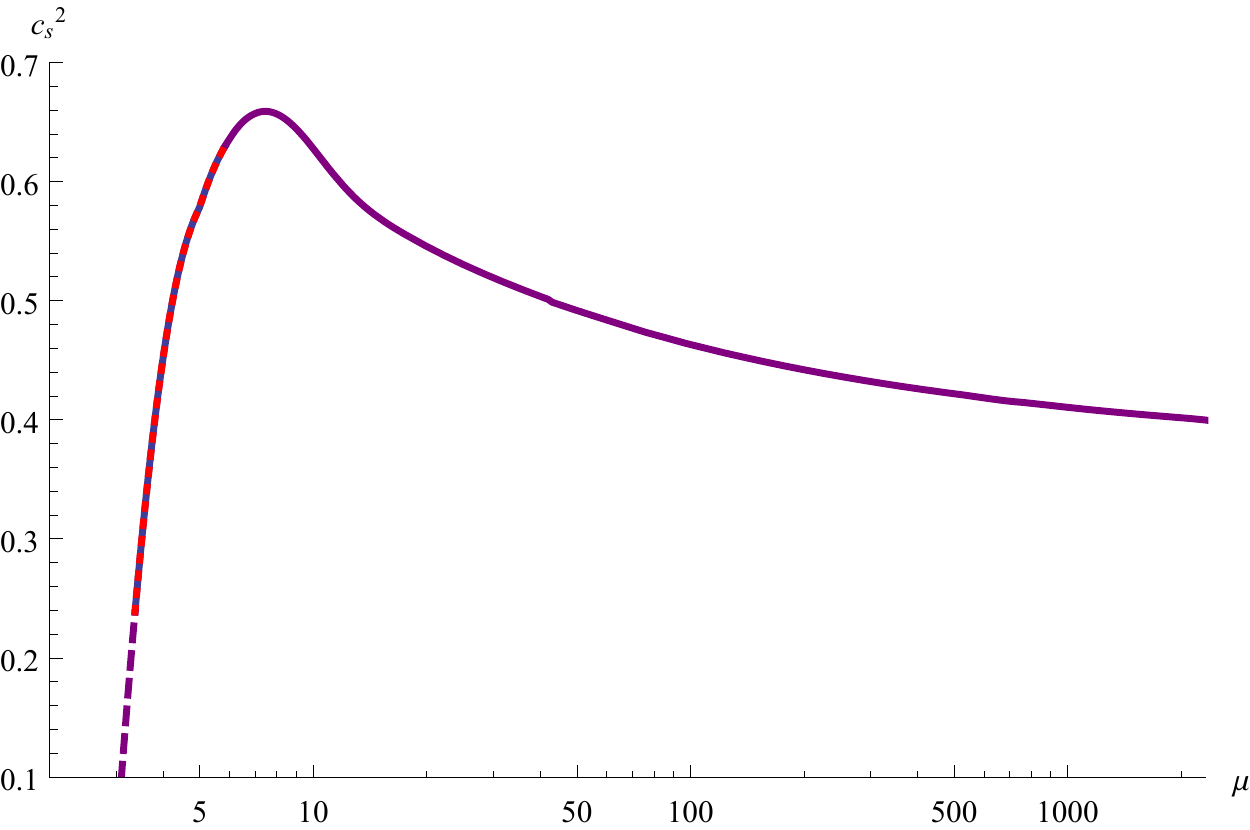} \includegraphics[width=0.49\linewidth]{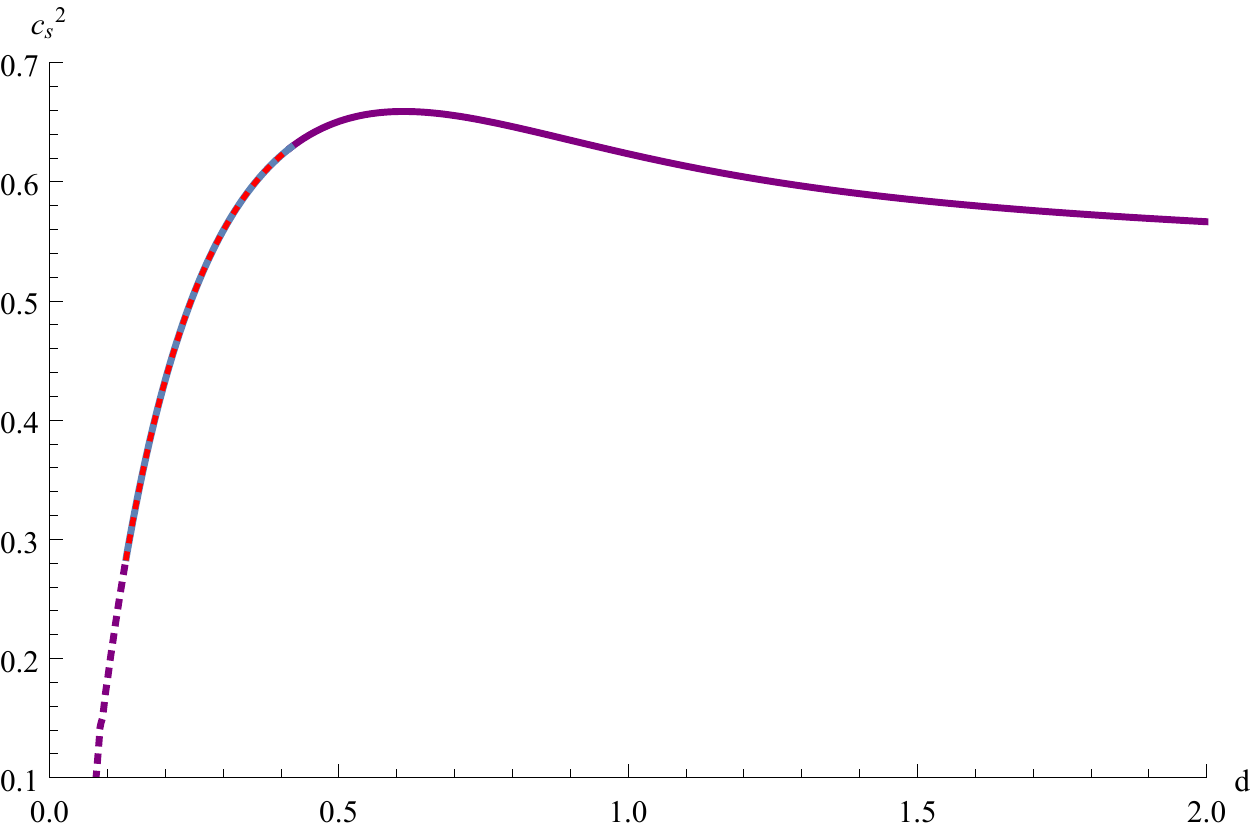}
    \caption{The squared speed of sound in the baryonic phase as a function of the chemical potential $\mu$ (left) and density $d$ (right) in units where $L=1$. The dashed sections represent unstable branches, never realized since at those chemical potentials the mesonic phase is energetically favored. The colors correspond (as will throughout all this article) to the two fit choices we employed, that we call fit A (blue) and fit B (red) in the next section. Where the curves coincide for the two fits, we draw a single purple line. The saturation density is approximately at $d_A\simeq 0.2$ for the fit A, and at $d_B\simeq 0.39$ for the fit B. The behavior is not monotonic and it lies above the ``sound barrier'' value of $1/3$, reaching a maximum close to double this value, enabling the possibility of highly stiff equations of state.}
    \label{fig:cs2}
\end{figure}

In early studies of holographic models, it was conjectured that the conformal value
of the speed of sound presents an upper bound \cite{Cherman:2009tw} for such models, but this was disproved later
by holographic models at finite density \cite{Hoyos:2016cob,Ecker:2017fyh,BitaghsirFadafan:2018uzs,Ishii:2019gta,BitaghsirFadafan:2019ofb,Jokela:2020piw,BitaghsirFadafan:2020otb}.\footnote{The Witten-Sakai-Sugimoto model also provides a counterexample \cite{Kulaxizi:2008jx,BitaghsirFadafan:2018uzs,Kovensky:2021kzl}, but does not respect the requirement in \cite{Cherman:2009tw} of a four-dimensional conformal fixed point in the UV.} In fact, the comparatively simple hard-wall AdS/QCD model that we are studying here also provides such a counterexample, which makes it possible to satisfy the empirical evidence from neutron stars that call for a speed of sound above the conformal value \cite{Bedaque:2014sqa,Kojo:2014rca,Tews:2018kmu}.

In the $T=0$ limit, the squared speed of sound can be obtained from the simplified formula:
\beq
c_s^2= \frac{d}{\mu}\left(\frac{\partial \mu}{\partial d} \right).
\eeq
In Fig.~(\ref{fig:cs2}) we plot the squared speed of sound in the baryonic phase with the parameter choices described in the previous section. As can be seen the speed of sound rapidly increases above the sound barrier of $c_s^2=1/3$, to reach almost double this value, then slowly decreases. This fits well the expectations of \cite{McLerran:2018hbz} for a quarkyonic nature of nuclear matter, which in our model is in fact continuously connected to the baryonic popcorn phase. If the latter is interpreted as a quarkyonic phase \cite{Kaplunovsky:2012gb}, nuclear matter in our model may be viewed as always having a quarkyonic nature.
However, also in the Witten-Sakai-Sugimoto model, where the quarkyonic phase obtained in \cite{Kovensky:2020xif} is separated from the nuclear matter phase, a speed of sound peaking to
values above the conformal one has been found in the ordinary nuclear matter phase \cite{BitaghsirFadafan:2018uzs} (albeit not as high as in our model).

Our numerical analysis indicates that for large $\mu$, the squared speed of sound decreases  continuously without falling below $1/3$. 
In Appendix \ref{app:soundspeed} we argue that asymptotically the conformal value is reached, in contrast to the Witten-Sakai-Sugimoto model, where the asymptotic value of the squared speed of sound is given by $2/5$ \cite{Kulaxizi:2008jx}.
At any rate, eventually perturbative QCD should take over, where the conformal
value is reached from below \cite{Kurkela:2009gj}.

	\subsection{Above the critical temperature}
	When $z_h$ reaches $z_{IR}$, all boundary conditions that encode spontaneous chiral symmetry breaking remain suddenly hidden beyond a black hole horizon, effectively keeping chiral symmetry restored in the theory for any value of $\mu$. Note that an explicit breaking (absent in this work) would be introduced via a quark mass which appears in the UV boundary condition for the scalar, so the explicit breaking cannot be undone (as expected).
	From these considerations follows that we can always assume the following boundary conditions for the scalar:
	\beq
	\omega_0(z_{h})=\omega_0(0)=0,
	\eeq
	and since its equation of motion is homogeneous, we can always trivially solve it for
	\beq
	\omega_0(z)=0.
	\eeq
	The presence of the black hole horizon forces us to choose Dirichlet boundary conditions for the fields $\widehat{L}_0,\widehat{R}_0$ in the infrared: the usual recipe is to set $\widehat{A}_0(z_{IR})=0$; however, we immediately see that this choice is not unique, since the value of the gauge field is not gauge invariant. We could set this value to any constant, modifying the interpretation of the value $\widehat{A}_0(z_{UV})$ as the chemical potential. To avoid such redefinition, we will choose:
	\beq
	\widehat{L}_0(z_{h})=\widehat{R}_0(z_{h})=0\qquad ;\qquad \widehat{L}_0(0)=\widehat{R}_0(0)=\mu.
	\eeq
Note that now the value of $\mu$ will determine the value of the derivative $\widehat{a}_0'(z)$, so it will enter the dynamics of the fields.

Starting from the easier scenario of $d=0$ (expected to hold for low chemical potential) we can see how a quark density arises: the equation of motion for $\widehat{a}_0$ in integrated form reads
	\beq
	-2M_5a(z)\widehat{a}'_0(z)=k,
	\eeq
	 with $k$ being an integration constant. 
	 If we were to impose the usual boundary condition $\widehat{a}_0'(z_{IR})=0$, then we would conclude that $k$ has to vanish for every temperature. However, the IR boundary is now hidden behind the horizon, where we have to impose a new boundary condition.
	 If we take as the boundary condition the vanishing of the field on the horizon (while retaining the UV boundary condition fixing $\mu$), then we end up with
	 \beq 
	 \widehat{a}_0(z) &=& -\frac{z^2}{4M_5}k + \mu,\\
	 k&=& \frac{4M_5}{z_h^2}\mu = 4M_5\pi^2 T^2 \mu.
	 \eeq
	 We can see from the holographic dictionary that $k$ is then associated to the baryon density, a role played before by the baryonic matter density from $H(z_{IR})$: in this phase we have no baryon matter explicitly built into the fields (in fact, this density does not contribute to the topological charge), so this contribution can be interpreted as coming from free quarks -- hence we identify the matter in this phase with a quark-gluon plasma.
	 This construction opens up a possibility for the presence of a quarkyonic phase along the lines of \cite{Kovensky:2020xif}, where in the top-down model of Witten-Sakai-Sugimoto, baryons are included forming a symmetrical layer far from the branes' joining point. For us, this amounts to relaxing the assumption that solitons are located on the IR brane (that is inaccessible in the chirally restored phase) and include both their contribution and that of free quarks in the equation of motion for the field $\widehat{a}_0$.
	 
	 \begin{figure}
    \centering
    \includegraphics[width=0.45\linewidth]{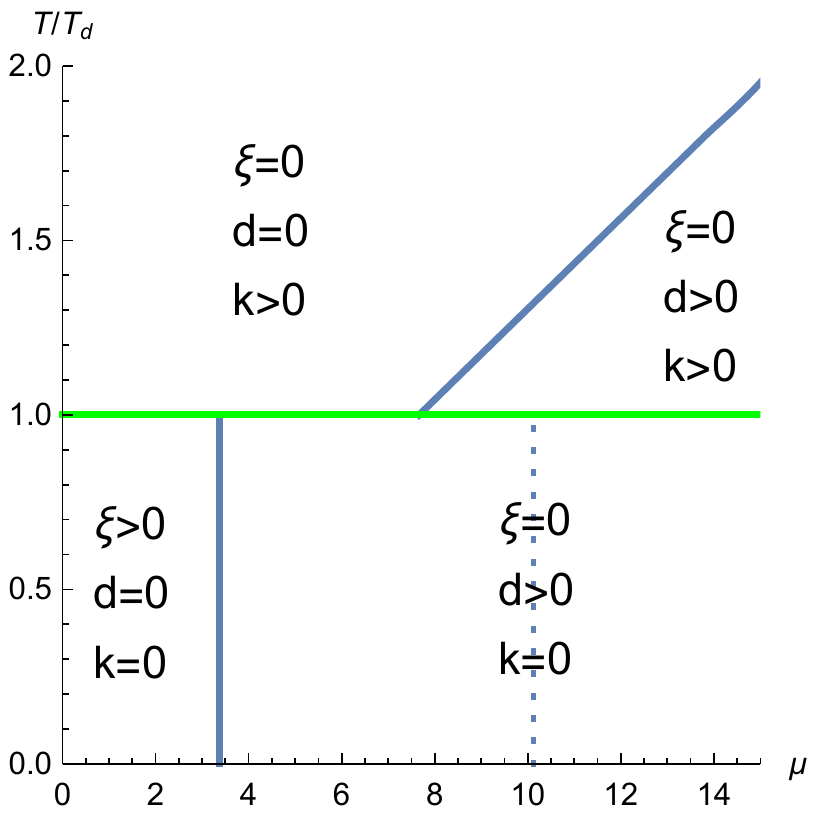}
    \caption{Phase diagram in the case of holographic QCD with a single IR wall ($z_0=z_{IR}$) with $\mu$ in units of $L^{-1}$; $\xi$ is the order parameter
    of chiral symmetry, $d$ and $k$ are baryon and deconfined quark densities, respectively. The vertical line for $T/T_d<1$ represents the first order transition to nuclear matter and chiral restoration, while the dotted line marks a crossover to a regime similar to the baryonic popcorn phase of \cite{Kaplunovsky:2012gb}. For $T/T_d>1$ and sufficiently large $\mu$ there is
    a phase of coexistence of deconfined quarks and baryons.}
    \label{fig:Phasediag0}
\end{figure}

	 \begin{figure}
    \centering
    \includegraphics[width=0.49\linewidth]{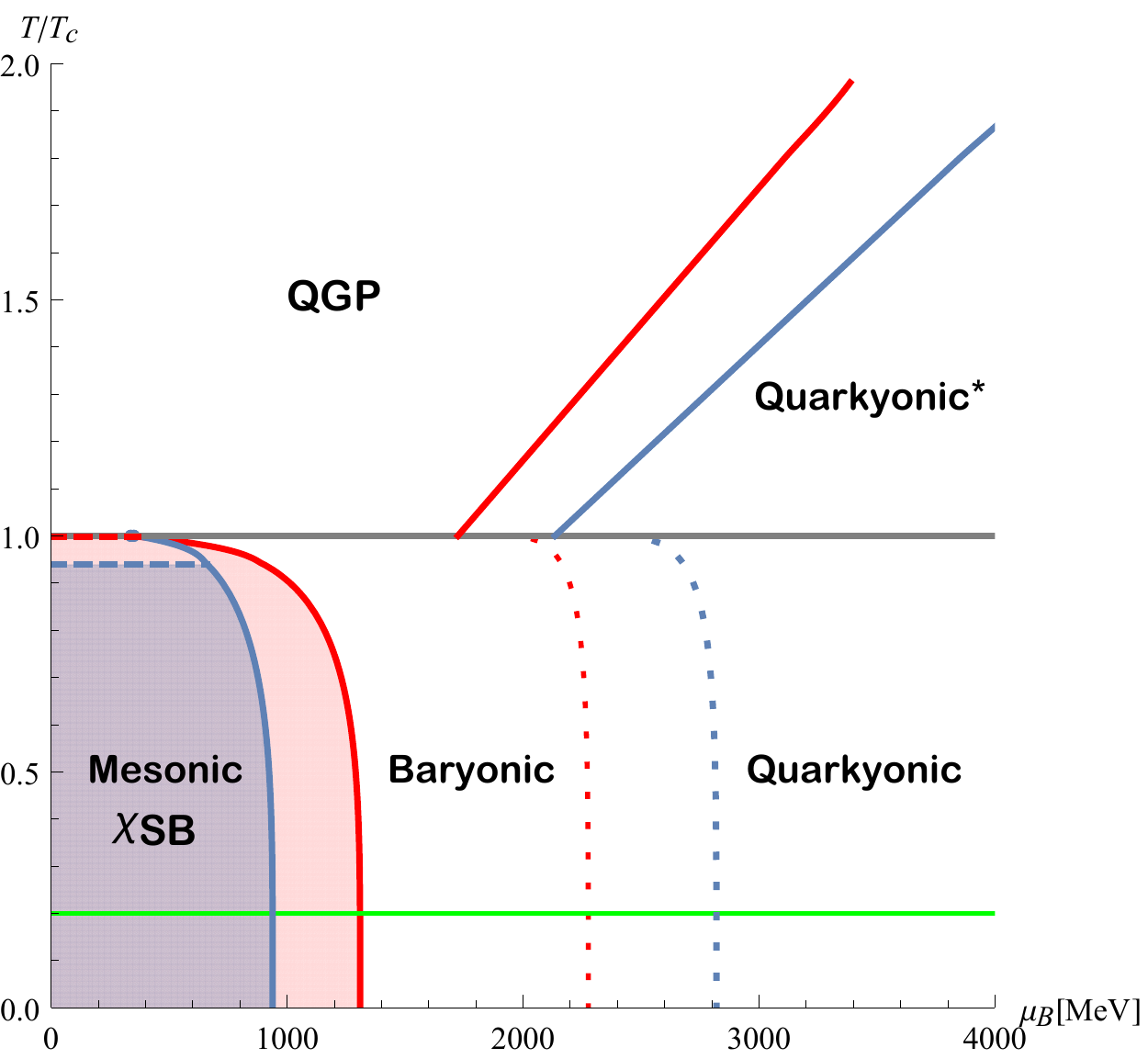}
     \includegraphics[width=0.49\linewidth]{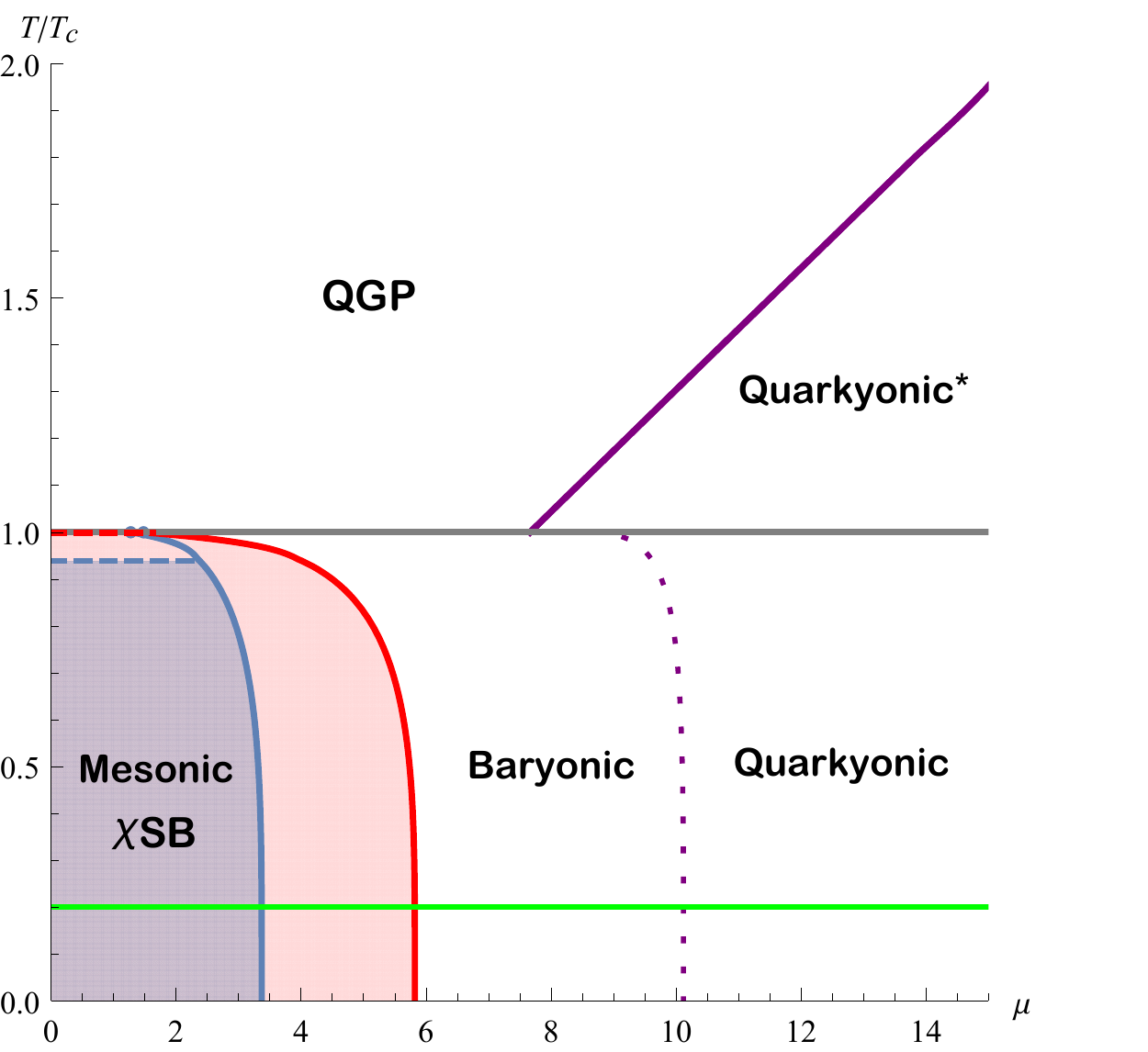}
    \caption{Phase diagrams of holographic QCD from the double hard-wall model (right panel with $\mu$ in units of $L^{-1}$, left panel in physical units). The green line represents a(n arbitrary in our setup) deconfining transition, below which every transition line is perfectly vertical (there is no temperature dependence). The blue and red colors represent respectively two choices of parameter sets: the blue one has parameters fitted to reproduce the correct chemical potential at the baryonic onset and the correct baryon mass from the hard-wall soliton (see App.~\ref{app:baryon}), while the red one has parameters chosen to have the equation of state lie predominantly in a phenomenologically relevant region, so as to obtain more realistic neutron stars.
    The solid curved lines for $T/T_c<1$ represent the corresponding first order transitions to nuclear matter and chiral restoration, while the dotted lines mark crossovers to a regime similar to the baryonic popcorn phase of \cite{Kaplunovsky:2012gb} that is identified with a quarkyonic phase.
    The horizontal dashed red and blue lines represent a second-order transition to chiral restoration within the mesonic phase, with the colored regions identifying phases with broken chiral symmetry. Above the critical temperature, a phase of quark-gluon plasma exists, while for large enough chemical potentials, and as a function of temperature, baryons also appear in this phase, realizing a quarkyonic phase of a fundamentally different nature than the one below $T_c$ with coexistence of deconfined quarks and baryons.}
    \label{fig:Phasediag1}
\end{figure}
 To do so, we turn on the $SU(2)$ fields with boundary conditions (i.e.~adding ``hair'' to the AdS black hole)
	 \beq
	 H(0)=0\qquad ;\qquad H(z_h)= (4\pi^2 d)^{\frac{1}{3}},
	 \eeq
	 and we solve again numerically the coupled system of $H,\widehat{a}_0$, choosing $d$ that minimizes the grand canonical potential. 
	 The numerical evaluation leads to the conclusion that a phase with both baryons and free quarks arises at high chemical potential for every temperature above $T_c$, with the transition curve being almost a straight line in the $\mu$-$T$ plane (see Figs.~\ref{fig:Phasediag0} and \ref{fig:Phasediag1}). The transition is of first order, and the baryonic matter appears always with its distribution peaked at a finite distance from the horizon, signaling a quarkyonic nature of this phase, which is marked ``quarkyonic$^*$'' in Figs.~\ref{fig:Phasediag0} and \ref{fig:Phasediag1}.\footnote{The results above are obtained in the absence of isospin chemical potential: for the derivation of a phase diagram within a holographic bottom-up model which includes effects of a finite isospin chemical potential, see \cite{Cao:2020ske}.}
	 
	 While Fig.~\ref{fig:Phasediag0} shows the qualitative structure in the
	 single-wall case ($z_0=z_{IR}$), where there is no temperature dependence
	 for $T<T_d$, Fig.~\ref{fig:Phasediag1} shows the numerical results obtained
	 for one arbitrary choice of $T_d<T_c$ and the resulting temperature dependence
	 of the otherwise straight lines separating mesonic and baryonic phases and the baryonic popcorn crossover.
	 The two sets of choices for the free parameters we employed here (and that we will employ in the next section) are the following:
	 
	 \beq\label{fitA}
	\text{Fit A: } && L^{-1}=186{\rm MeV}\qquad ;\qquad \lambda =  2\times 10^{-3}\qquad ;\qquad \xi_0= 1.05,\\
	\label{fitB}
\text{Fit B: }	&& L^{-1}=150{\rm MeV}\qquad ;\qquad \lambda \xi_0^4 = 1.024,
	 \eeq
	 where the fit A indicates one that correctly reproduces the baryon mass and the critical chemical potential for the baryon onset, and the fit B is chosen as an example of a fit that produces an equation of state which lies as much as possible within the constraints by observational data from neutron stars. For fit A the values of $\xi_0$ and $\lambda$ are independently relevant, since the combination $\frac{1}{2}\lambda \xi_0^4$ is the depth of the potential well set by the infrared potential (and as such it governs the value of the baryon onset), while only $\xi_0$ enters the mass of the single baryon calculated in Appendix \ref{app:baryon}. For fit B instead only the value $\lambda \xi_0^4$ enters the calculations through the value of the baryon onset, since the baryonic phase always sets $\xi=0$.

	 It should be noted that the phase diagram that we have obtained here does not include
	 backreactions of the flavor physics on the geometry, which is always AdS. These backreactions
	 will certainly become increasingly important as the baryon density is increased such that
	 the phase transition at $T/T_c=1$ acquires a nontrivial dependence on $\mu$. The expectation is that $T_c(\mu)$ decreases with increasing $\mu$. If so, the phase diagram
	 that we have obtained in Fig.~\ref{fig:Phasediag1} in terms of $T/T_c$ vs $\mu$ could be mapped to one in terms of $T$ vs $\mu$ where the horizontal lines bend down at large $\mu$,
	 and if its structure remains intact it could be interpreted as predicting a
	 phase transition inside the baryonic phase, where the latter does not need to extend to high absolute temperatures but could in fact be restricted to the low-temperature domain.
	 
\section{Neutron stars}\label{sec:neutron}

In the following we shall take the above results for the low-temperature equation of state
at face value and consider the consequences for a toy neutron star formed by the corresponding baryonic matter. Unlike the holographic models reviewed in \cite{Jarvinen:2021jbd,Hoyos:2021uff} we refrain from matching to realistic
equation of states for conventional nuclear matter at moderate
densities, thereby completely neglecting any effects from the crust of
a neutron star.
The modeling of a phenomenologically viable neutron star
crust cannot be done purely within the homogeneous Ansatz and will
require additional input, such as the physics of individual baryons
and the presence of leptons, or simply a hybrid matching with other
phenomenological equations of state. 
Also,
the nontrivial phase diagram that we have obtained above for temperatures above the deconfinement temperature will play no role here.

\subsection{Tolman-Oppenheimer-Volkov equations}
Neutron stars are described by the Tolman-Oppenheimer-Volkov (TOV) equations, which read:
\beq
\frac{dP}{dr}&=& -G(\varepsilon+ P)\frac{m+4\pi r^3 P}{r(r-2Gm)},\\
\frac{dm}{dr}&=&4\pi r^2 \varepsilon.
\eeq
The equations are solved by using as a boundary condition $P(r=0)=P_0$ for a range of values of $P_0$, and the radius of the neutron star obtained, is defined as the value of $R$ for which $P(R)=0$. Different values of $P_0$ will produce results of $R,m$ that define a curve in the $m$-$R$ plot to be compared with observational data.
The TOV equation is dependent on the equation of state of the baryonic matter in the model, and since our quantities at equilibrium and $T=0$ are solely dependent on $\mu$, we effectively solve for $\mu$ by minimizing the grand potential to obtain $d(\mu)$, and feeding it to $P,\varepsilon$ so as to obtain $P(\mu,d(\mu)), \varepsilon(\mu,d(\mu))$.

Since we work in the grand canonical ensemble, we use the holographic dictionary as before to identify the grand potential with (minus) the on-shell action. Then we can use the relations:
\beq
PV &=& -\Omega  = V \int dz \mathcal{L}^{\rm on-shell} ,\\
\varepsilon &=& \frac{E}{V} = -P + \mu \frac{\partial P}{\partial \mu} = - P + N_c \frac{\mu}{2}d.
\eeq
In agreement with the identification of the grand potential as $\Omega = E-\tilde{\mu} N $: note that we have an additional factor of $1/2$ that consistently stems from the definition of $\mu$ we employed.
	 \begin{figure}[!htp]
    \centering
    \includegraphics[width=0.52\linewidth]{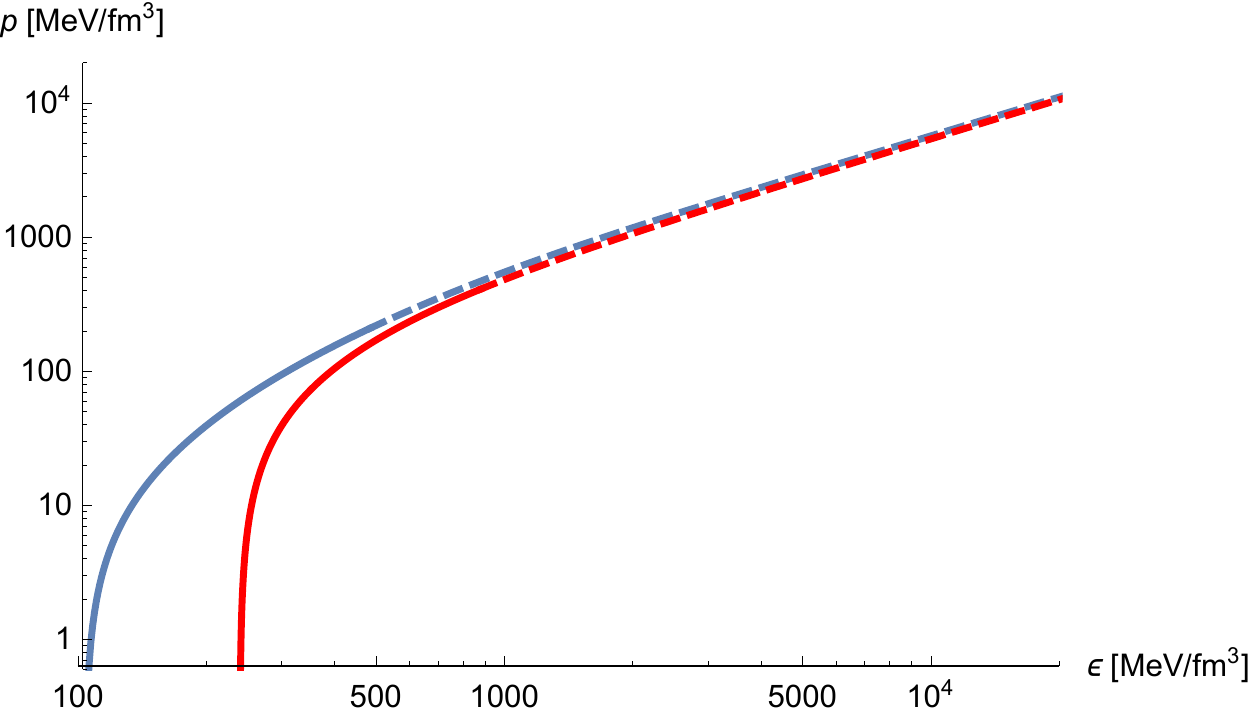}
     \includegraphics[width=0.47\linewidth]{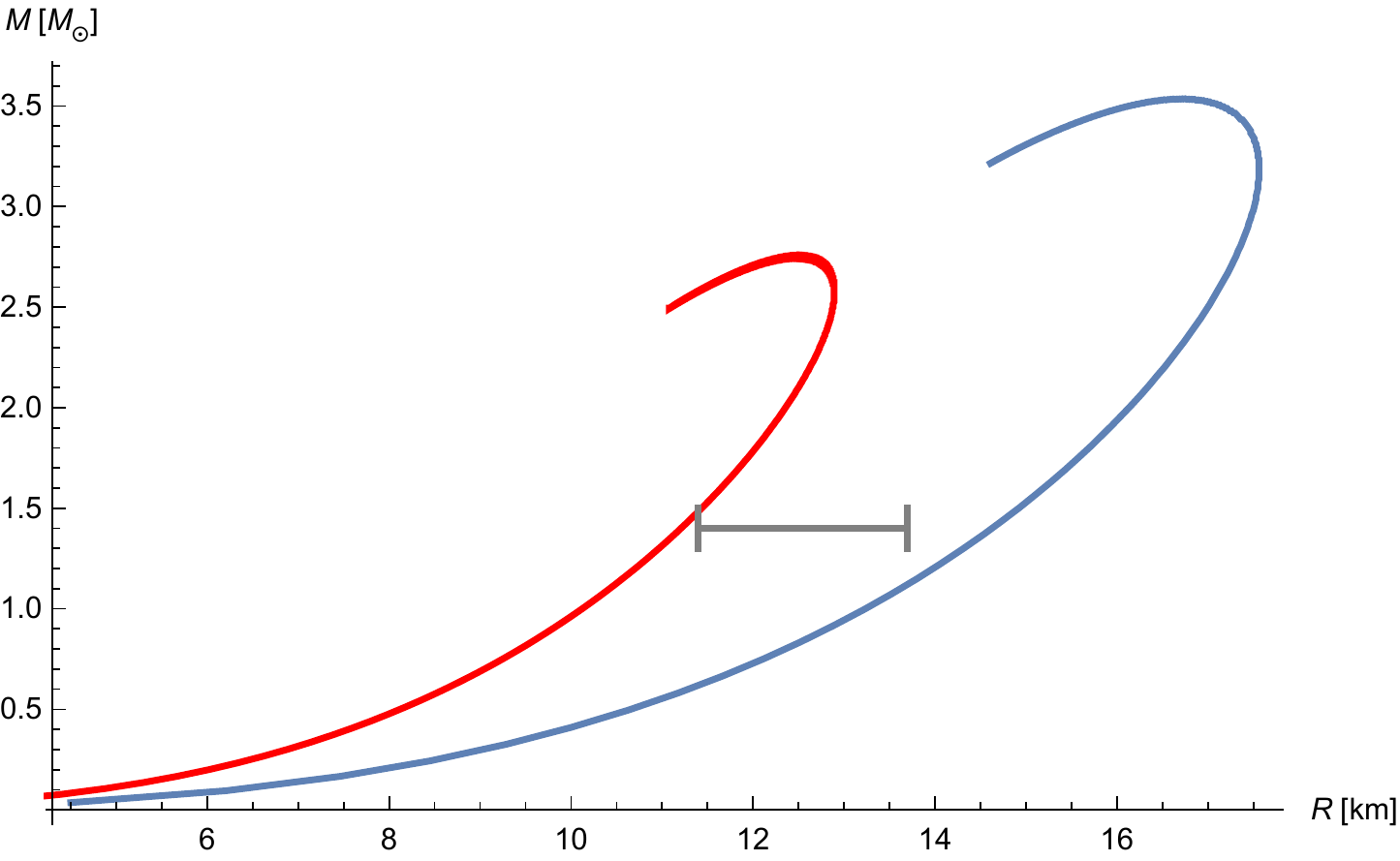}
    \caption{Equations of state (left) and radius-mass relations for neutron stars with two choices of parameters. The dashed sections of the equations of state indicate values of central pressure that lead to unstable neutron stars. In the mass-radius plot, these stars lie beyond the turning point (corresponding to the maximum mass) of the respective curves. For the blue lines (fit A), we fitted the baryonic mass and the chemical potential at baryon onset to the phenomenological values, leaving us with one free parameter which is then chosen to reproduce the best possible mass-radius relation, which in this case is quite far from what is expected, leading to quite heavy stars. For the red plots (fit B) all parameters are instead chosen to obtain a good compromise between highest mass, radius for stars of $1.4M_\odot$ (whose most recent measurement by NICER is represented by the gray interval), and tidal deformability (see Fig. \ref{tidals}). We see that in both cases the measured radius is not completely compatible with our predictions: however, the effect of a crust is expected to increase the radius of neutron stars, refining the precision of the ``phenomenological'' set of parameters.}
    \label{eosMRplots}
\end{figure}
With all these identifications we can plot the $P$-$\varepsilon$ curve that defines our equation of state, and the $m$-$R$ curve that arises from the solutions to the TOV equations.
The identification of $P$ and $\varepsilon$ with the dual holographic quantities can also be done by constructing the stress-energy tensor $T^i_j$ and reading off $T^0_0$ and $T^i_i$ from there. In general, the hydrodynamic pressure found this way is different from the thermodynamical one, but they coincide at equilibrium, as we have verified numerically (see Appendix \ref{app:stress-energy} for this derivation of $P,\varepsilon$). 

In Fig.~\ref{eosMRplots} we plot the equation of state and the mass-radius relations for both the  fit of the parameters according to (\ref{fitA}), and an alternate scenario (\ref{fitB}) where we do not fit any observable quantity, but we choose parameters such that the equation of state falls as much as possible within the window allowed by observational data, and such as to obtain more realistic mass-radius relations.
With fit A (\ref{fitA}), we find that the speed of sound inside stable neutron stars never reaches the maximum allowed by the equation of state, peaking instead at a value of $c_s^2=0.645$ at the center of the most massive stable star, while the polytropic index $\gamma=d\ln P/d\ln\varepsilon$ reaches a minimum value of $\gamma=1.5$; with fit B (\ref{fitB}) instead the allowed speed of sound values extend beyond the peak in the curve $c_s^2(\mu)$, so that the maximum value reached corresponds to the global maximum $c_s^2=0.659$, while the polytropic index inside stable neutron stars reaches a lower minimum value of $\gamma = 1.31$. 
Similarly to the results obtained in Ref.~\cite{Kovensky:2021kzl} with the Witten-Sakai-Sugimoto model, we thus find that at the center of neutron stars
$\gamma$ drops below the value 1.75 used in Ref.~\cite{Annala:2019puf} as criterion for the presence of quark matter, while within the holographic model the nuclear matter phase persists.

\begin{figure}[!htp]
    \centering
    \includegraphics[width=0.6\linewidth]{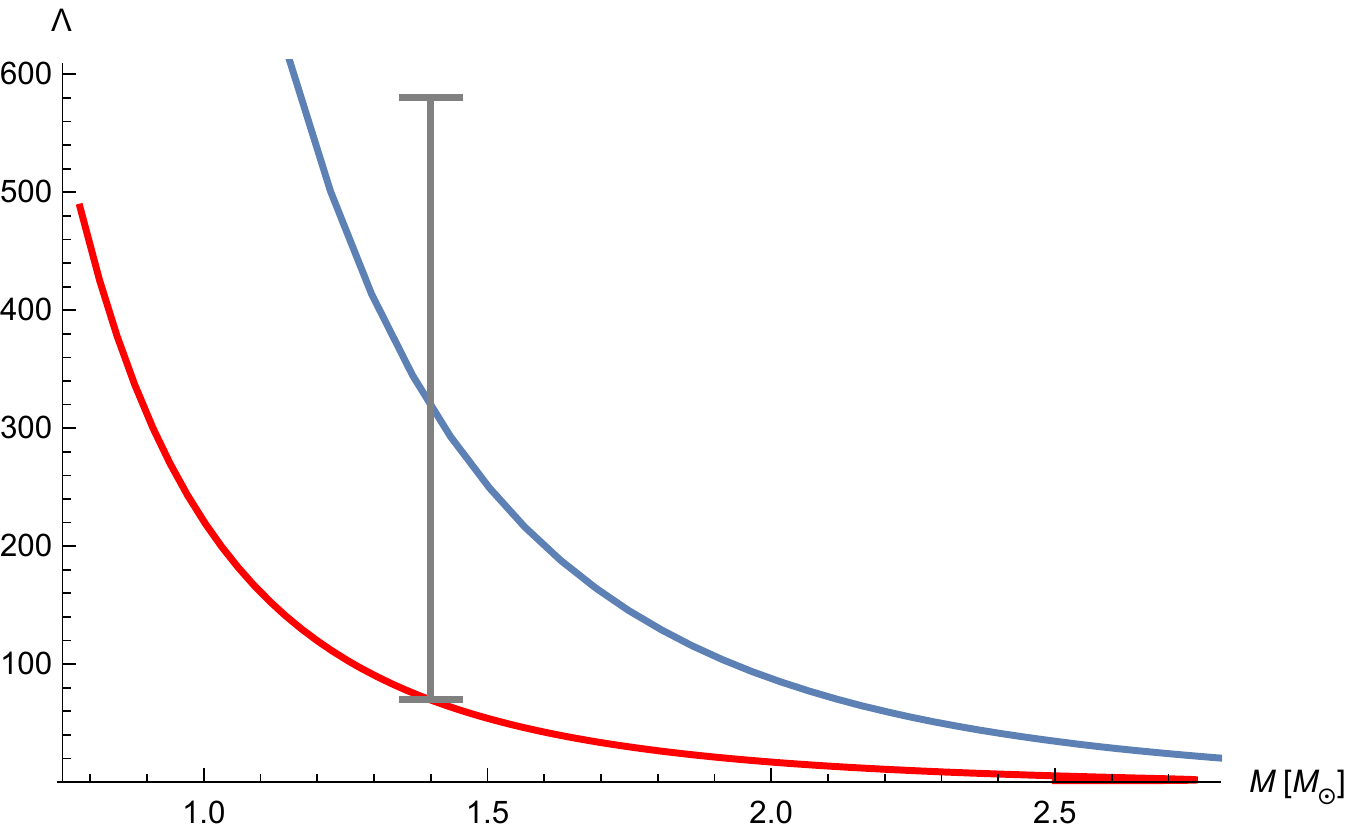}
    \caption{Tidal deformabilities as a function of star mass in the two parameter choices considered. Color coding of the plots is as in Fig.~\ref{eosMRplots}. We see that both fits lead to acceptable results, with the ``phenomenological'' one being extremely close to the lower end of the allowed region. Effects of a crust are expected to increase tidal deformabilities, pushing this last fit to more realistic predictions.}
    \label{tidals}
\end{figure}
Another important property of neutron stars is the tidal deformability: in Fig.~\ref{tidals} we plot the tidal deformability $\Lambda$ as a function of the star's mass: we see that by choosing the more phenomenologically accurate equation of state, we obtain the least likely tidal deformability for a star with mass of $1.4M_\odot$, at the lower boundary of the allowed values. The equation of state obtained by fitting the baryon mass and onset instead produces a tidal deformability well within the experimental bounds.

\subsection{Neutron star mergers and gravitational waves}

We will now use the equation of state (EoS) found in the previous sections and tested by the TOV equations in a full numerical gravity simulation, giving rise to a gravitational wave signal, in principle detectable at the LIGO and VIRGO experiments.

The setup that we will consider is a binary of two neutron stars, each of mass $1.4$ solar masses ($M_\odot$) with an initial separation of 45 km.
The initial configuration is made with \texttt{LORENE} \cite{Lorene} using first \texttt{init\_bin} and then \texttt{coal} with suitable initial enthalpies (large enough for creating an initial configuration of the desired mass.
We perform the numerical gravity simulations using \texttt{WhiskyTHC} \cite{Radice:2012cu,Radice:2013hxh,Radice:2013xpa} using the method of lines with a third order Runge-Kutta (RK3) evolution, a hybrid EoS mode in the ``thorn'' \texttt{EOS\_Thermal} with the holographic EoS input as the cold EoS and the temperature dependence given by a simple Gamma law with coefficient $\Gamma_{\rm th}=1.75$ \cite{Ecker:2019xrw}: this accounts for shock-heating effects during the merger \cite{Bauswein:2010dn} by adding a thermal component to the equation of state with $P_{\rm th}=\Gamma_{\rm th}\rho(\varepsilon_{\rm total}-\varepsilon_c)=\Gamma_{\rm th}\rho\varepsilon_{\rm th}$ where $\rho=m_b d$ is the baryon mass density, $m_b$ is the baryon rest mass and $d$ is the number density. The total pressure is then $P$ from the cold equation of state plus the thermal component $P_{\rm th}$, i.e.~$P_{\rm total}=P+P_{\rm th}$.
\begin{figure}[!htp]
\begin{center}
\mbox{\includegraphics[width=0.33\linewidth]{{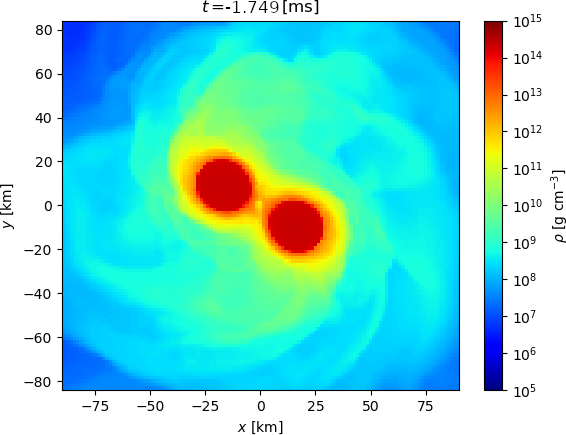}}
\includegraphics[width=0.33\linewidth]{{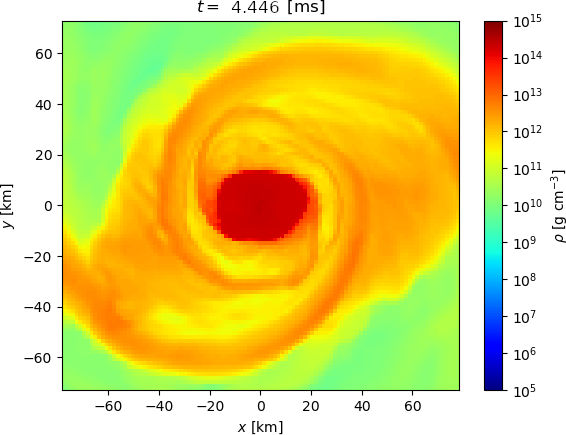}}
\includegraphics[width=0.33\linewidth]{{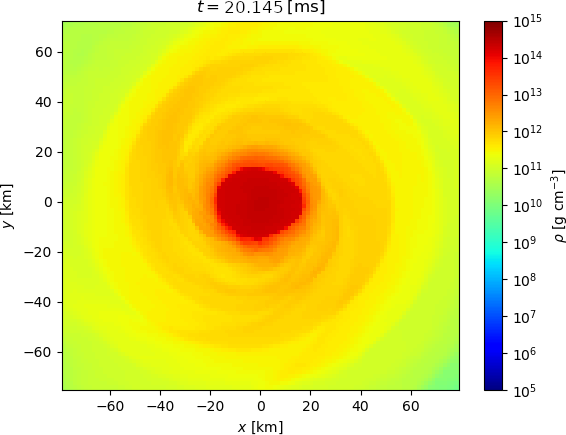}}}\\
\hbox{\hspace{1.67cm}\includegraphics[width=0.9\linewidth]{{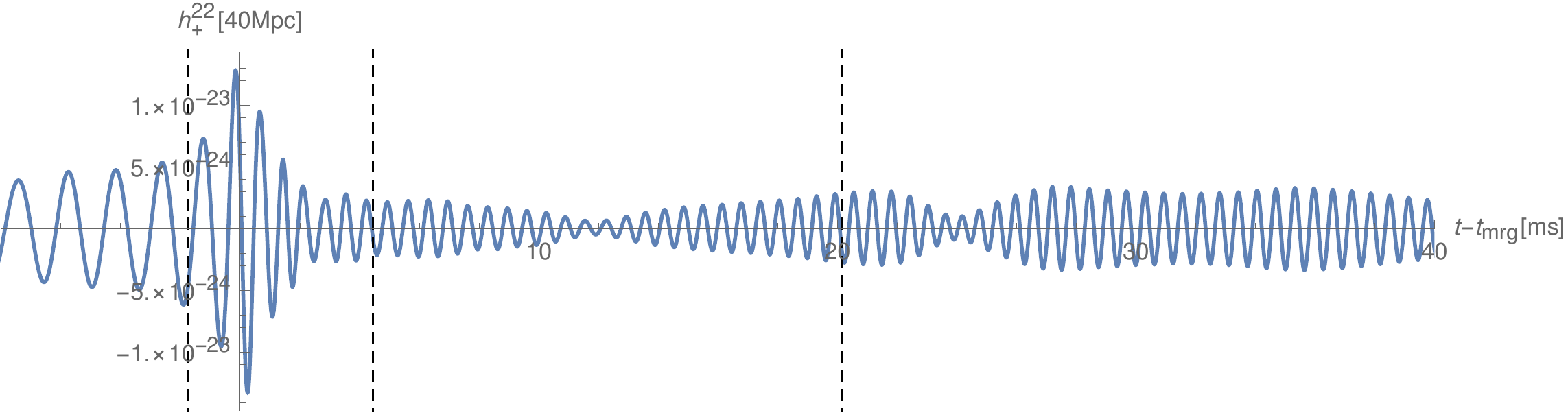}}}
\includegraphics[width=0.87\linewidth]{{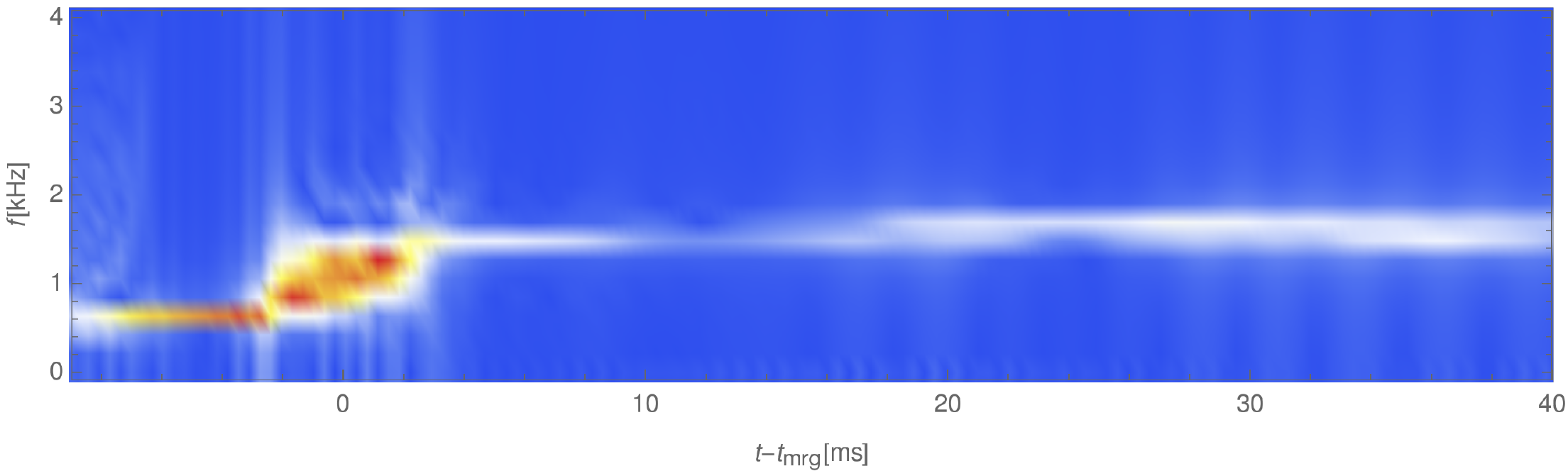}}
\end{center}
\caption{The gravitational wave signal of a two $1.4M_\odot$ neutron stars merger, in the model fitted to the baryon mass and baryon onset, measured from the fourth Weyl scalar $\psi_4$ in terms of the spherical harmonic decomposition into the function $h_+^{22}$, extrapolated to 40Mpc. The bottom row shows the frequencies of the signal as a function of time using bins of roughly 2 ms, whereas the top row of figures shows snapshots of the hydrodynamic general relativistic numerical simulation. }
\label{fig:signal1}
\end{figure}
\begin{figure}[!htp]
\begin{center}
\mbox{\includegraphics[width=0.33\linewidth]{{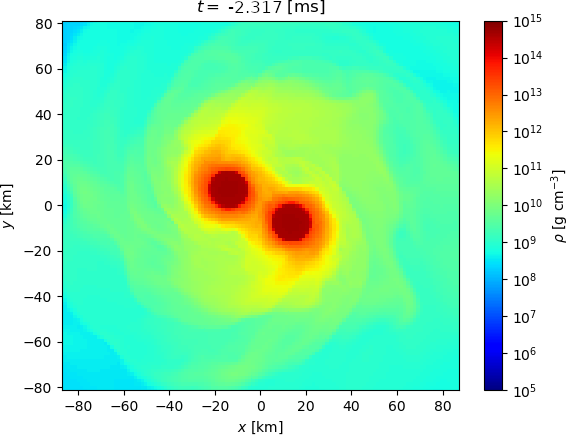}}
\includegraphics[width=0.33\linewidth]{{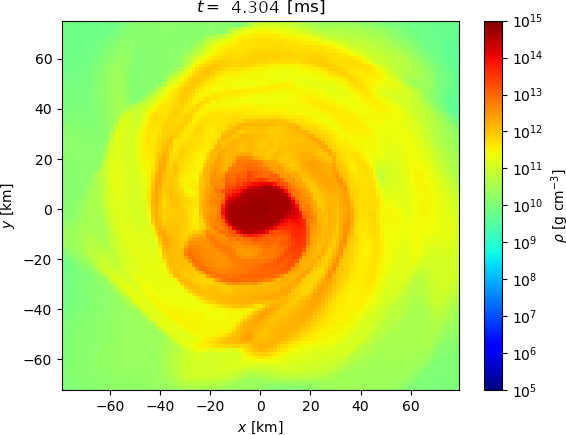}}
\includegraphics[width=0.33\linewidth]{{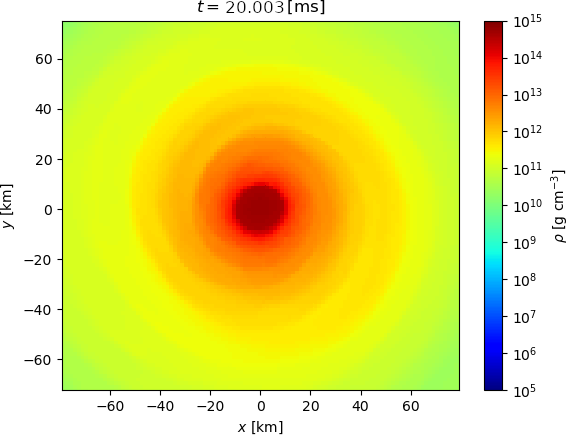}}}\\
\hbox{\hspace{1.67cm}\includegraphics[width=0.9\linewidth]{{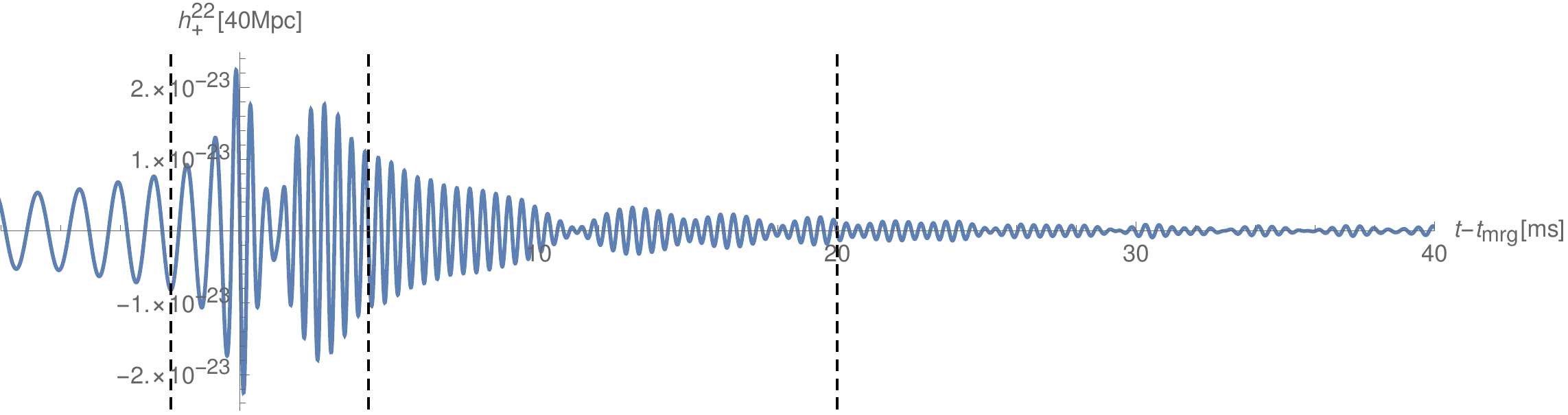}}}
\includegraphics[width=0.87\linewidth]{{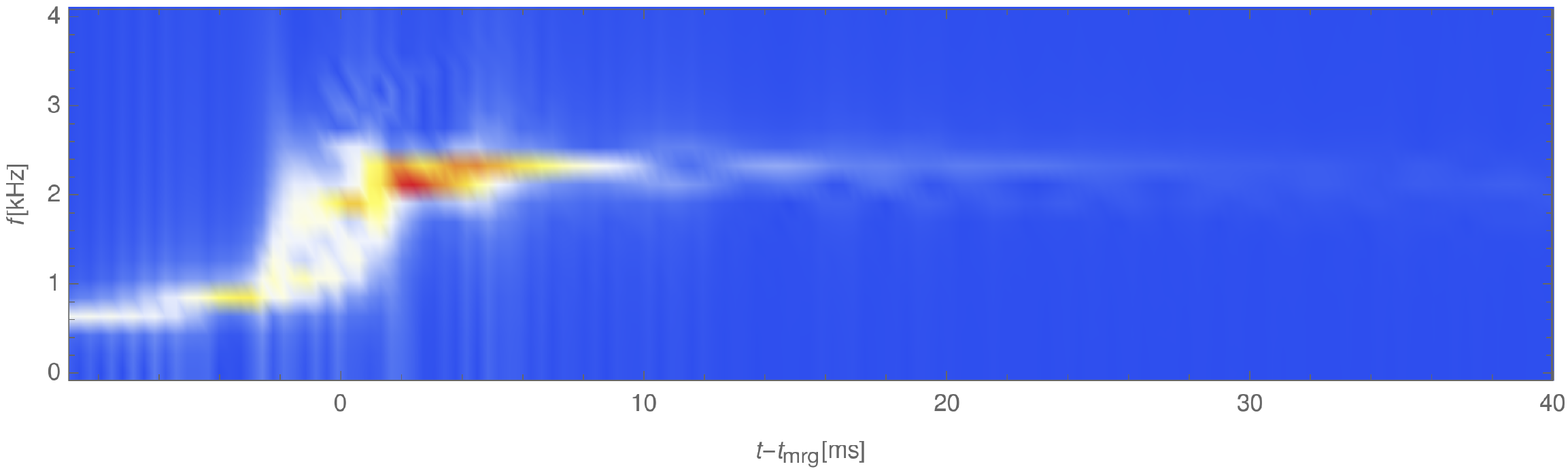}}
\end{center}
\caption{The gravitational wave signal of a two $1.4M_\odot$ neutron stars merger, in the model with the choice of parameters made to optimize the equation of state, measured from the fourth Weyl scalar $\psi_4$ in terms of the spherical harmonic decomposition into the function $h_+^{22}$, extrapolated to 40Mpc. The bottom row shows the frequencies of the signal as a function of time using bins of roughly 2 ms, whereas the top row of figures shows snapshots of the hydrodynamic general relativistic numerical simulation. }
\label{fig:signal2}
\end{figure}
The result for the choice of parameters fitted to the baryon mass and baryon onset is shown in Fig.~\ref{fig:signal1}, whereas the result for the ``phenomenological'' parameter choice is shown in Fig.~\ref{fig:signal2}, both for $1.4M_\odot+1.4M_\odot$ neutron stars as initial condition.

\section{Conclusions and outlook}\label{sec:conclusions}
We studied the phase diagram of a holographic ``hard-wall'' model of QCD, identifying baryonic and quarkyonic phase transitions for arbitrary temperatures, with a nontrivial temperature dependence arising from a generalization to separate infrared walls for gluonic and quark degrees of freedom. To do so, we employed the approximation of homogeneous baryonic matter, which holds true for sufficiently high densities, while we expect corrections to this picture at low densities to arise by treating baryons as individual topological solitons. At high densities we found an analogue of the baryonic popcorn transition observed before in \cite{Kaplunovsky:2012gb} in the Witten-Sakai-Sugimoto model, which we interpret as a continuous crossover to a quarkyonic phase. Another quarkyonic phase emerged at high densities above the critical temperature for chiral symmetry restoration, which could be taken as an indication that in a more refined treatment with backreacted geometry and $\mu$-dependent critical temperature there is a larger quarkyonic regime with embedded phase transition.

In the low-temperature phases, we have evaluated the resulting equation of state
with a view towards modeling cold nuclear matter at high densities.
Independently of the fit chosen for the free parameters in our model, the resulting speed of sound at zero temperature in this baryonic matter turned out to be rather high, quickly rising above the conformal value of $c_s^2=1/3$ and reaching a maximum of approximately $c_s^2=0.66$ before monotonically decreasing with further increases of the chemical potential. A steep rise of
the speed of sound to values of order one has been argued in \cite{McLerran:2018hbz} to
be associated with a quarkyonic nature of baryonic matter, 
which is consistent with the continuous connection between the baryonic and quarkyonic (baryonic popcorn) phases in our model.

We used the equation of state of the model to solve the TOV equations and build neutron stars: to do so, we employed two different sets of parameter choices, one that reproduces the correct baryon mass and the correct baryon onset, while the other gives a good compromise for the observable properties of neutron stars.
The resulting neutron stars turn out to be quite compact, with maximum masses higher than what is expected, with the radius for stars of mass $1.4 M_\odot$ being slightly outside (smaller than) the allowed interval, and with tidal deformability for the same star exactly on the lower bound of the allowed interval. However, we have to bear in mind an important limitation in our setup: the homogeneous Ansatz is expected to only describe accurately the core of neutron stars, while we do not have a description for a phase with more diluted solitons, which would provide the stars with a softer crust. The effects of the presence of a crust are expected to be an increase in the radius of stars (at fixed mass), and an increase in the tidal deformability. To put this intuition to the test, we extended our holographic equation of state with the tabulated ``SLy4'' one \cite{Douchin:2001sv} and built again neutron stars with the phenomenological set of parameters. As a result, the radius of the $1.4 M_\odot$ star increased to over $12.3$ km, while its tidal deformability increased over the allowed interval, to a value of about $775$. This result opens up the possibility to obtain a better phenomenological fit by slightly changing parameters in order to obtain a lower maximum mass, which would in turn also lower the tidal deformability and the radius. We leave the modeling of a suitable crust phase for neutron stars within the same holographic model, and thus a more refined computation of neutron stars' properties, for a future work.

\subsection*{Acknowledgments}

We would like to thank Andreas Schmitt for useful discussions and Christian Ecker and Matti J\"arvinen for correspondence and help with the \texttt{WhiskyTHC} code.
The work of L.~B.~is supported by the National Natural Science Foundation of China (Grant No.~12150410316). This work was initiated while L.~B.~was at TU Wien, supported by a grant from "Fondazione Angelo Della Riccia", which supports Italian young researchers for working abroad.
S.~B.~G.~thanks the Outstanding Talent Program of Henan University for
partial support.
S. B. G. thanks the Ministry of Education of Henan Province for partial support.
The work of S.~B.~G.~is supported by the National Natural Science
Foundation of China (Grants No.~11675223 and No.~12071111).
J.~L.~has been supported by the FWF doctoral program Particles \& Interactions, project no.~W1252-N27, and FWF Project no.~P33655.

\begin{appendices}
	
	\section{The speed of sound in the limit of large densities}\label{app:soundspeed}
	In this appendix we will consider taking the limit of large chemical potential, which corresponds to large densities. We will do the considerations for $T=0$ and in the baryonic phase, so $\xi=0$. The equation of motion \eqref{eqomega} for $\omega_0$ is trivial for $\xi=0$ and hence $\omega_0=0$ everywhere.
	$H$ has to obey the boundary conditions \eqref{eq:HUV} and \eqref{eq:HIR}, which for large $d$ becomes very energetically expensive. We conjecture, based on numerical experience, that the limit of the function $H(z)$ tends to a step function for $\mu\to\infty$:
	\beq
	\lim_{\mu\to\infty}H(z) = (4\pi^2d)^{\frac13}\Theta_H(z),
	\eeq
	and for large $\mu$, we have that $d\propto\mu^\alpha$ with $\alpha\geq2$.
	Integrating the Chern-Simons term by parts, we have the contributing terms in the Lagrangian:
	\beq
	\mathcal{L} = -3M_5a(z)H^4 
	+\frac{N_c}{8\pi^2}\partial_z(\widehat{a}_0 H^3).
	\eeq
	Integrating over $z$ we obtain
	\beq
	L = -3M_5(4\pi^2d)^{\frac43}\mathcal{I} + \frac{N_c\mu d}{2},
	\eeq
	where we have defined the regularized integral
	\beq
	\mathcal{I} := \int a(z)\Theta_H(z-\epsilon(d))dz = -\log\epsilon(d) > 0,
	\eeq
	where $\epsilon(d)$ parametrizes our ignorance about how the true solution approaches the step function in the limit of $d\to\infty$.
	Minimizing the energy with respect to $d$ yields
	\begin{align}
	-\frac{\partial L}{\partial d}&=
	4M_5(4\pi^2)^{\frac43}\left(d^{\frac13}\mathcal{I} - d^{\frac43}\frac{\epsilon'(d)}{\epsilon(d)}\right)
	- \frac{N_c\mu}{2}\nonumber\\
	&=4M_5(4\pi^2)^{\frac43}(\mathcal{I} + \gamma) d^{\frac13}
	- \frac{N_c\mu}{2}
	=0,
	\end{align}
	where in the second line, we have assumed a power-law behavior of $\epsilon(d)\propto d^{-\gamma}$.
	The solution to the above equation in terms of $\mu^3$ is
	\beq
	\mu^3 = \frac{2(\mathcal{I}+\gamma)^3M_5^3(4\pi)^8d}{N_c^3},
	\eeq
	where $\mathcal{I}$ is logarithmically dependent on $d$.
	In the limit $d\to\infty$ the speed of sound becomes:
	\beq
	\lim_{d\to\infty} c_s^2 = \lim_{d\to\infty} \frac{d}{\mu}\left(\frac{\partial d}{\partial\mu}\right)^{-1}
	=\frac13,
	\eeq
	and the conformal value of the speed of sound squared is thus obtained.
	
	\section{The baryon as an instanton}\label{app:baryon}
	In this appendix, we will calculate the mass of the baryon using the Ansatz
	\begin{align}
	R_i(\bx,z) &= A_1(r,z)\frac{\bx\cdot\btau x^i}{2r^2}
	+\frac{1}{2r^2}(1-\phi_1(r,z))\tau^a\epsilon_{a i k}x^k
	+\frac{1}{2r}\phi_2(r,z)\left(\frac{\bx\cdot\btau x^i}{r^2} - \tau^i\right),\\
	L_i(\bx,z) &= -A_1(r,z)\frac{\bx\cdot\btau x^i}{2r^2}
	+\frac{1}{2r^2}(1-\phi_1(r,z))\tau^a\epsilon_{a i k}x^k
	-\frac{1}{2r}\phi_2(r,z)\left(\frac{\bx\cdot\btau x^i}{r^2} - \tau^i\right),\\
	R_5(\bx,z) &= -L_5(\bx,z) = \frac{1}{2r}A_2(r,z)\bx\cdot\btau,\\
	R_0(\bx,z) &= L_0(\bx,z) = \frac{s(r,z)}{2r}\mathds{1},\\
	\Phi(\bx,z) &= \lambda_2\frac{\mathds{1}}{2} + i\lambda_1\frac{\bx\cdot\btau}{2r},
	\end{align}
	for a single instanton, following \cite{Pomarol:2007kr,Pomarol:2008aa,Domenech:2010aq}, giving rise to the reduced energy functionals
	\begin{align}
	E_g &= 8\pi M_5\int_0^\infty dr\int_{z_{UV}}^{z_{IR}} dz
	\bigg[
	a(z)\left(|D_{\bar{\mu}}\phi|^2 + \frac{r^2}{4}A_{\bar{\mu}\bar{\nu}}^2
	+\frac{(1-|\phi|^2)^2}{2r^2}
	-\frac12(\partial_{\bar{\mu}}s)^2
	\right)\nonumber\\&\phantom{=8\pi M_5\int_0^\infty dr\int_{z_{UV}}^{z_{IR}} dz\bigg[}
	-\frac{N_c}{32\pi^2M_5}\frac{s}{r}\epsilon^{\bar{\mu}\bar{\nu}}
	\left(\partial_\mu\left(-i\phi^* D_{\bar{\nu}}\phi + i\phi(D_{\bar{\nu}}\phi)^*\right) + A_{\bar{\mu}\bar{\nu}}\right)
	\bigg],\label{eq:E_g_baryon}\\
	E_\Phi &= 8\pi M_5\int_0^\infty dr\int_{z_{UV}}^{z_{IR}}dz
	\left[a^3(z)\left(
	\frac{r^2}{4}|D_{\bar{\mu}}\lambda|^2
	-\frac18(\phi\lambda^*-\lambda\phi^*)^2\right)
	+a^5(z)\frac{r^2}{4}M_{\bulk}^2|\lambda|^2
	\right],\label{eq:E_Phi_baryon}
	\end{align}
	where we have defined the complex fields
	\beq
	\phi = \phi_1 + i\phi_2, \qquad
	\lambda = \lambda_1 + i\lambda_2,
	\eeq
	the covariant derivative $D_{\bar{\mu}}=\partial_{\bar{\mu}}-iA_{\bar{\mu}}$, the abelian field strength $A_{\bar{\mu}\bar{\nu}}=\partial_{\bar{\mu}}A_{\bar{\nu}}-\partial_{\bar{\nu}}A_{\bar{\mu}}$ and finally, the barred indices, $\bar{\mu},\bar{\nu}$ run over $r,z$ with Euclidean metric.
	
	The equations of motion relevant for the instanton are
	\begin{align}
	    D_{\bar{\mu}}(a(z)D_{\bar{\mu}}\phi)
	    +\frac{a(z)}{r^2}\phi(1-|\phi|^2)
	    +\frac{a^3(z)}{4}\lambda(\lambda\phi^*-\phi\lambda^*)
	    +\frac{iN_c}{16\pi^2M_5}\epsilon^{\bar{\mu}\bar{\nu}}
	    \partial_{\bar{\mu}}\left(\frac{s}{r}\right)D_{\bar{\nu}}\phi = 0,\\
	\partial_{\bar{\mu}}(r^2a(z)A_{\bar{\mu}\bar{\nu}})
	-a(z)\big(i\phi^*D_{\bar{\nu}} - i\phi(D_{\bar{\nu}}\phi)^*\big)
	-\frac{ia^3(z)r^2}{4}\big(\lambda^*D_{\bar{\nu}}\lambda - (D_{\bar{\nu}}\lambda)^*\lambda\big)\qquad\qquad\nonumber\\
	\mathop+\frac{N_c}{16\pi^2M_5}\epsilon^{\bar{\mu}\bar{\nu}}
	\partial_{\bar{\mu}}\left(\frac{s}{r}\right)(|\phi|^2-1) = 0,\\
	\partial_{\bar{\mu}}(a(z)\partial_{\bar{\mu}}s)
	-\frac{N_c}{32\pi^2 M_5 r}\epsilon^{\bar{\mu}\bar{\nu}}
	\left[\partial_{\bar{\mu}}\left(-i\phi^*D_{\bar{\nu}}\phi +i\phi(D_{\bar{\nu}}\phi)^*\right)
	+A_{\bar{\mu}\bar{\nu}}\right] = 0,\\
	D_{\bar{\mu}}(r^2a^3(z)D_{\bar{\mu}}\lambda)
	-a^3(z)\phi(\lambda\phi^*-\phi\lambda^*)
	-a^5(z)r^2M_{\bulk}^2\lambda = 0,
	\end{align}
	the Lorentz gauge condition is
	\beq
	\partial_{\bar{\mu}}A_{\bar{\mu}} = 0,
	\eeq
	the boundary condition in the infrared ($z=z_{IR}$) are
	\beq
	\phi_1 = 0, \quad
	\p_z\phi_2 = 0,\quad
	A_r = 0,\quad
	\partial_z A_z = 0,\quad
	\partial_z s = 0, \quad
	\lambda = -2i\xi,
	\eeq
	and in the ultraviolet ($z=z_{UV}$) they are
	\beq
	\phi = -i,\quad
	A_r = 0,\quad
	\partial_z A_z = 0,\quad
	s = 0,\quad
	\lambda = 0,
	\eeq
	whereas at spatial infinity, the boundary condition for the single instanton are
	\beq
	\phi = -ie^{\frac{i\pi(z-z_{UV})}{z_{IR}-z_{UV}}},\quad
	\partial_r A_r = 0,\quad
	A_z = \frac{\pi}{z_{IR}-z_{UV}},\quad
	s = 0,\quad
	\lambda = 2i\xi z^3 e^{\frac{i\pi(z-z_{UV})}{z_{IR}-z_{UV}}},
	\eeq
	and at the origin we have
	\beq
	\phi_1 = 0,\quad
	\phi_2 = -1,\quad
	\partial_r A_r = 0,\quad
	A_z = 0,\quad
	s = 0,\quad
	\lambda_1 = 0,\quad
	\partial_r\lambda_2 = 0.
	\eeq
	By using suitable initial conditions and a numerical scheme sometimes called arrested Newton flow, we solve the PDEs numerically and integrate the energy functionals \eqref{eq:E_g_baryon} and \eqref{eq:E_Phi_baryon} to find the energy of the instanton, which is identified with the mass of the baryon (nucleon).
	\begin{figure}[!htp]
	\begin{center}
	    \includegraphics[width=\linewidth]{{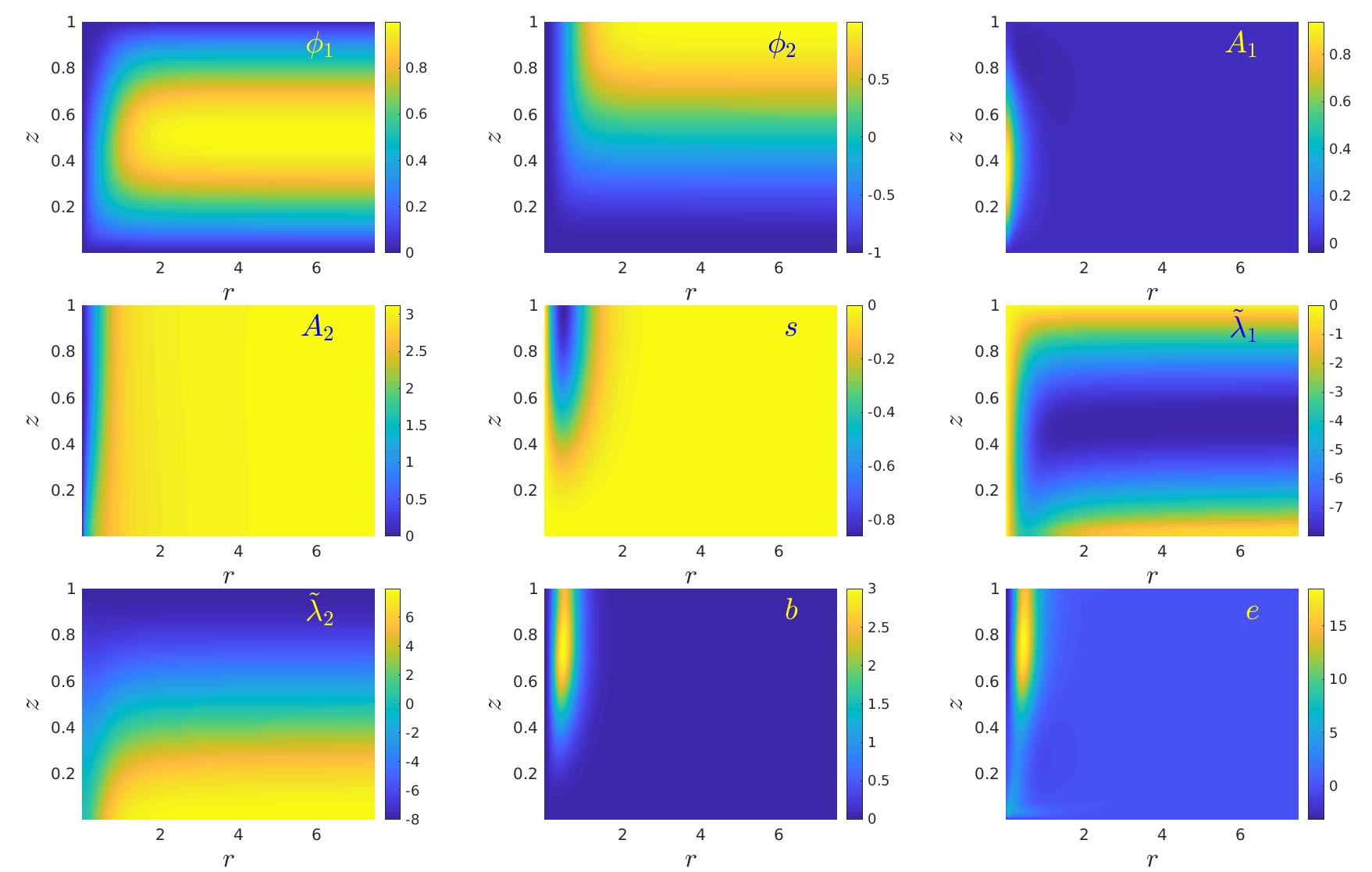}}
	    \caption{Instanton solution in the bulk for $\xi=4$, $M_5=\frac{N_c}{12\pi^2}$, $N_c=3$ and $M_{\bulk}^2=-3$.}
	\label{fig:instanton4}
	\end{center}
	\end{figure}
	In Fig.~\ref{fig:instanton4} we show an example of the instanton solution in the bulk for $\xi=4$, $M_5=\frac{N_c}{12\pi^2}$, $N_c=3$ and $M_{\bulk}^2=-3$.
	For numerical convenience, we have defined a new scalar field which is defined as being relative to the vacuum solution (up to a factor of $\xi$):
    \beq
	\lambda = \lambda_1+i\lambda_2 = z^3\tilde{\lambda}.
	\eeq
	\begin{figure}[!htp]
	\begin{center}
	\mbox{\subfloat[]{\includegraphics[width=0.49\linewidth]{{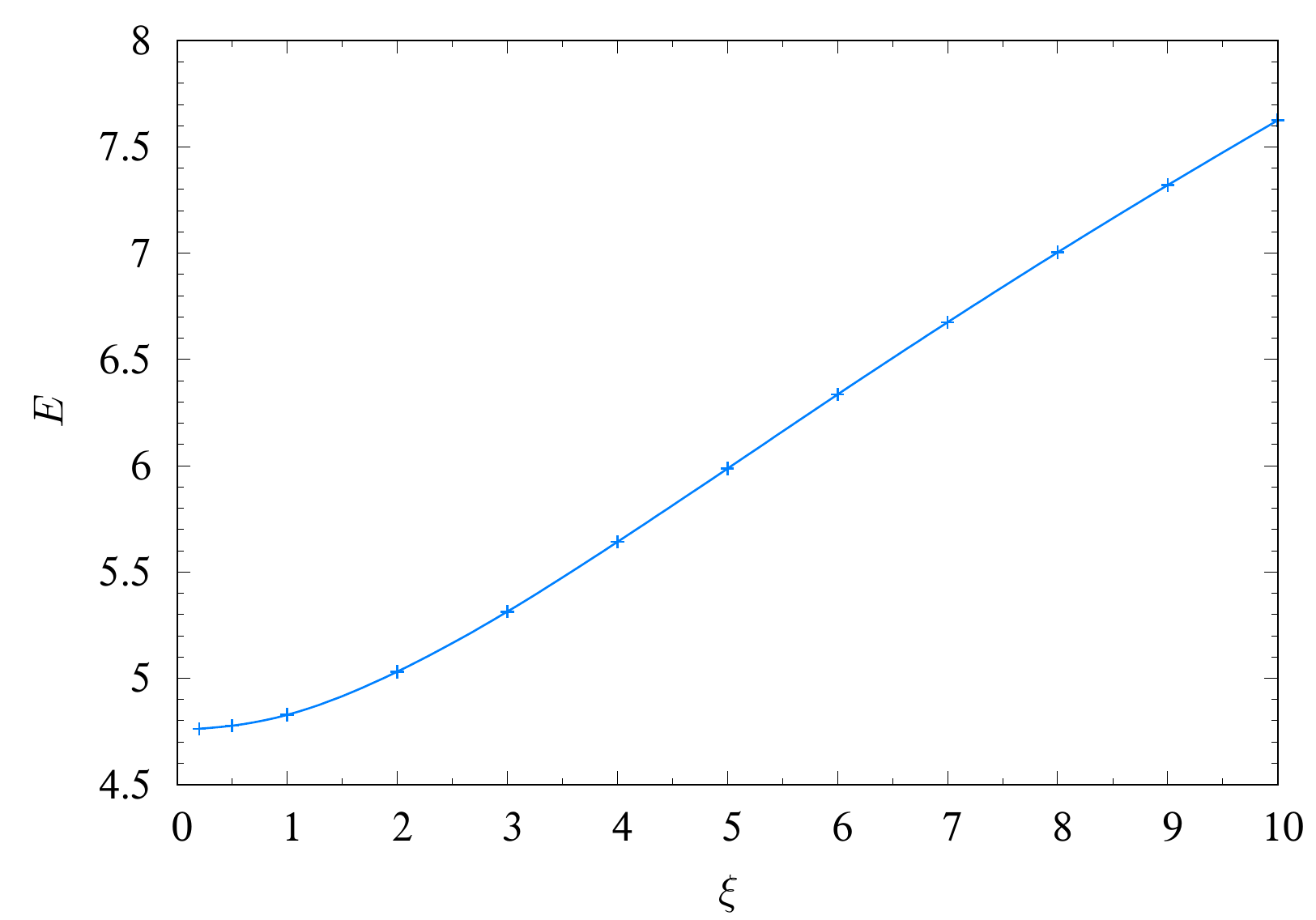}}}
	\subfloat[]{\includegraphics[width=0.49\linewidth]{{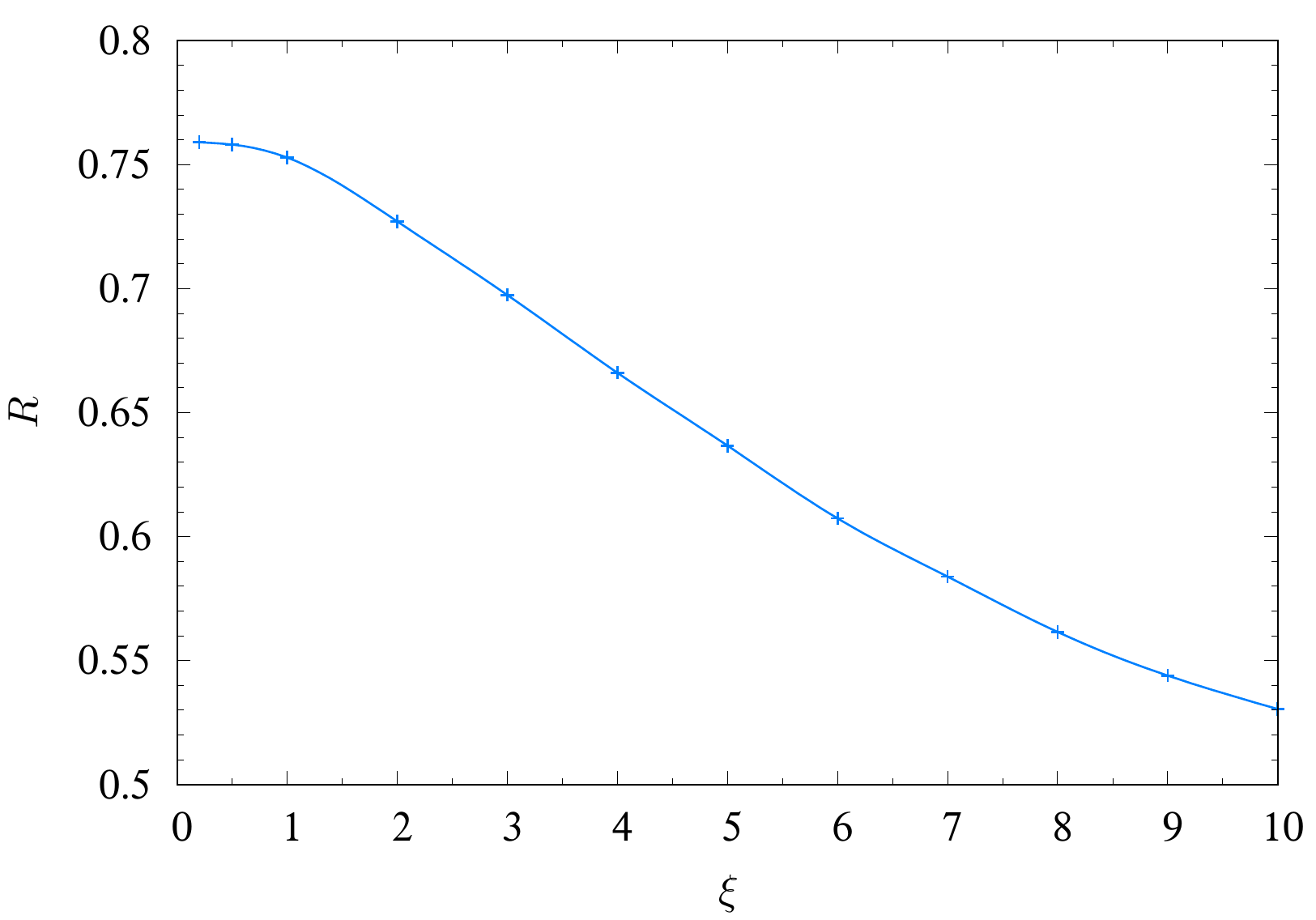}}}}
	    \caption{Energy (a) and radius (b) as functions of $\xi$. In this figure, we have taken the parameters of the model as $M_5=\frac{N_c}{12\pi^2}$, $N_c=3$ and $M_{\bulk}^2=-3$.}
	\label{fig:energy_radius}
	\end{center}
	\end{figure}
	Changing $\xi$ affects the energy and the size or ``radius'' of the instanton. The result for a range of $\xi$s is shown in Fig.~\ref{fig:energy_radius}.
	We use the data of Fig.~\ref{fig:energy_radius} to calibrate the model by fitting the energy of the instanton to the mass of the proton/neutron (we work in the limit of unbroken isospin):
	\beq
	L = \frac{E_{\rm instanton}}{M_{\rm proton}},
	\eeq
	where $L$ is the curvature scale in the metric as given in Eq.~\eqref{eq:metric}.
	
		\section{Boundary conditions and baryon number}\label{app:BC}
	
	Here we discuss our choice of boundary conditions for the homogeneous Ansatz. As noted in \cite{Domenech:2010aq}, a gauge transformation $g_{L,R}=\exp(i\alpha_{L,R})$ that does not reduce to unity at the UV brane correctly reproduces the QCD global anomaly, but also carries an infrared term which is not present in QCD and which should be canceled for the holographic model to be consistently describing QCD.
	
	The action is gauge invariant except for the Chern-Simons term, whose variation amounts to:
	\beq
	\Delta_\alpha S_{CS} = \frac{N_c}{24\pi^2}\left(\int_{UV}\left[\bar{\omega}_4^1(\alpha_L,l) - \bar{\omega}_4^1(\alpha_R,r)\right] -\int_{IR} \left[  \bar{\omega}_4^1(\alpha_L,L)-\bar{\omega}_4^1(\alpha_R,r) \right]  \right)  .
	\eeq
   We want to cancel the infrared term, which can be done by the following choice of boundary conditions:
   \beq
  ( L_{\mu}-R_{\mu})|_{z_{IR}} = 0\qquad ;\qquad(  L_{\mu z}+R_{\mu z})|_{z_{IR}}=0.
   \eeq
   The second one of these conditions is automatically satisfied by the homogeneous Ansatz, while the first is in general not satisfied for $\mu=i$:
   \beq\label{usualbound}
   ( L_{i}-R_{i})|_{z_{IR}} = - H(z_{IR}) \sigma^i = -(4\pi^2 d)^{1/3} \sigma^i.
   \eeq
   If we want to describe baryonic matter within this Ansatz, we have to give up on this choice of boundary condition, as it is incompatible with $H^3(z_{IR}) =4\pi^2 d$.
   However, this does not mean that the infrared variation of the Chern-Simons term cannot vanish. Let us take a look at $\bar{\omega}_4^1$;
   \beq\label{defomega}
   \bar{\omega}_4^1(\alpha,A) = \frac{1}{4}\widehat{\alpha} \big( d\widehat{A}\big)^2 +\frac{3}{2} \widehat{\alpha}  \tr F^2.
   \eeq
   The conditions (\ref{usualbound}) enforce the vanishing of the variation by cancellation between the $L$ and $R$ terms: another possibility is that they both vanish independently which is precisely what happens for the homogeneous Ansatz.
   The first term in Eq.~(\ref{defomega}) amounts to
  \beq
  \frac{1}{4}\widehat{\alpha} \big( d\widehat{A}\big)^2 =  \frac{1}{4}\widehat{\alpha} \widehat{F}_{\alpha_1\alpha_2}\widehat{F}_{\alpha_3\alpha_4} dx^{\alpha_1} \wedge dx^{\alpha_2} \wedge dx^{\alpha_3} \wedge dx^{\alpha_4}.
  \eeq
  Since the form lives on a four-dimensional surface orthogonal to $z$, then actually $\alpha_i = \mu_i$. But from the homogeneous Ansatz we know that the only dependence of the fields is on $z$, and only the $0$-th component of the abelian field is turned on, so that:
  \beq
  \widehat{F}_{\mu \nu} = 0,
  \eeq
  hence the first term vanishes.
  Similarly, the second term is proportional to:
  \beq
  \tr F^2\propto F_{\mu_1 \mu_2}^a F_{\mu_3 \mu_4}^a  dx^{\mu_1} \wedge dx^{\mu_2} \wedge dx^{\mu_3} \wedge dx^{\mu_4},
  \eeq
  and because of anti-symmetry of the wedge product it will select the only term with all indices different, which then vanishes because of the presence of $F_{0k}$. 
  \beq
  F_{ij}^a F_{0k}^a\epsilon^{ijk} = 0.
  \eeq
  In the end, each $\bar{\omega}_4^1(\alpha,A)$ independently vanishes and the unwanted anomaly is still canceled with our choice of boundary conditions in the homogeneous Ansatz.
		\section{Pressure and energy density from stress-energy tensor}\label{app:stress-energy}
Here we want to compute the pressure and the energy density of the baryonic phase, starting from the stress-energy tensor, verifying that the definitions match the relations with the grand-canonical potential. We start by recalling the definition of the stress-energy tensor $T_{\alpha\beta}$ of matter fields in general relativity:
\beq
T_{\alpha\beta}= \frac{2}{\sqrt{-g}}\frac{\partial\left(\sqrt{-g} \mathcal{L}_{\rm matter}\right)}{\partial g^{\alpha \beta}},
\eeq
where $\mathcal{L}_{\rm matter}$ indicates the Lagrangian density of matter fields, not including the $\sqrt{-g}$ factor (as it is part of the integration measure). 
In our model, there are two kinds of action terms for matter fields: the ones from $\mathcal{L}_g,\mathcal{L}_{\Phi}$ which depend on the metric, and the Chern-Simons one from $\mathcal{L}_{CS}$, which is topological and hence does not contribute to $T_{\alpha\beta}$ by definition (it can and will still contribute via the equations of motion).
In the baryonic phase, and in the chiral limit we are considering, the scalar field vanishes because of the minimum energy boundary condition $\xi=0$, hence $\mathcal{L}_\Phi$ also vanishes, so for us holds the following identification:
\beq
\mathcal{L}_{\rm matter}= -\frac{M_5}{2} \left[\Tr\left(L_{MN}L^{MN} \right)+\frac{1}{2}\widehat{L}_{MN}\widehat{L}^{MN} + \left\{R\leftrightarrow L\right\}\right],
\eeq
where now the indices are to be lowered with the full curved metric $g^{\alpha\beta}$, as opposed to with $\eta^{\alpha\beta}$ as was the case in Eq.~(\ref{Sgauge}).
To perform the computation it is useful to use the following form of the stress-energy tensor:
\beq
T_{\alpha\beta}=T_{\alpha\beta}^{(1)} + T_{\alpha\beta}^{(2)}
=2\frac{\partial \mathcal{L}_{\rm matter}}{\partial g^{\alpha \beta}}-g_{\alpha \beta}\mathcal{L}_{\rm matter}.
\label{eq:Tuseful}
\eeq
We want to compute the energy density $\mathcal{E}$ and the hydrodynamic pressure $P$ of the model:
\beq
\varepsilon = \frac{1}{V_3}\int d^3x dz \sqrt{-g}T^0_0\qquad;\qquad P= -\frac{1}{3 V_3}\int d^3x dz \sqrt{-g}T^j_j.
\eeq

Let us begin by computing the energy density, starting from the first term of Eq.~\eqref{eq:Tuseful}, remembering that with the homogeneous Ansatz, only the field strengths $L_{ij},L_{iz}$ and $\widehat{L}_{0z}$ are turned on:
\beq
T_{00}^{(1)}=2\frac{\partial \mathcal{L}_{\rm matter}}{\partial g^{00}}&=&-M_5\left( g^{zz}\widehat{L}_{0z}\widehat{L}_{0z}+g^{zz}\widehat{R}_{0z}\widehat{R}_{0z} \right)\nonumber\\
&=&2M_5 a^{-2}(z)\left(\partial_z \widehat{a}_0\right)^2.
\eeq
We now raise one index and perform the integration by parts:
\beq
\frac{1}{V_3}\int d^3x dz \sqrt{-g}T^{0(1)}_0 &=& 2M_5\int dz \sqrt{-g}g^{00}\left(\partial_z\widehat{a}_0\right)^2\nonumber\\
&=&2M_5\int dz\; a(z)\left(\partial_z\widehat{a}_0\right)^2\nonumber\\
&=&2M_5\big[a(z)\widehat{a}_0\partial_z\widehat{a}_0\big]^{1}_{0}-2M_5\int dz\; \partial_z\big(a(z)\partial_z\widehat{a}_0\big)\widehat{a}_0.
\eeq
The integral term can be immediately seen to give rise to $-L_{CS}$ after making use of the equation of motion for the field $\widehat{a}_0$. The boundary term vanishes in the infrared, but despite the warp factor approaching zero on the ultraviolet boundary, there also $\widehat{a}_0'$ vanishes. We can then make use of the ultraviolet expansion of $\widehat{a}_0$ to compute the contribution of this term:
\begin{align}
\widehat{a}_0(z\rightarrow0) = \mu-z^2\frac{N_c d}{8 M_5}+\ldots,\\
2M_5\left[a(z)\widehat{a}_0\partial_z\widehat{a}_0\right]^{1}_{0} = \frac{N_c}{2}\mu d.
\end{align}
The second term in the definition of the stress-energy tensor is proportional to $\mathcal{L}_{\rm matter}$, and since this differs from $\mathcal{L}_g$ only by a warp factor, it is easy to check that once we raise one index and include the factor of $\sqrt{-g}$ we obtain the simple contribution:
\beq
-\frac{1}{V_3}\int d^3x dz \sqrt{-g} g_0^0\mathcal{L}_{\rm matter}= -L_g,
\eeq
so that when we sum the two contributions and we use the definition for $\mathcal{E}$ we obtain
\beq
\varepsilon = -L_{CS} - L_g +\frac{N_c}{2}\mu d = -L^\mathrm{on-shell} +\mu_B d,
\eeq
where $L^\mathrm{on-shell}$ is the total on-shell Lagrangian integrated over $z$. 

We now turn to compute the hydrodynamic pressure: again we divide the computation into two parts, this time starting with the trivial one
\beq
T^{p(2)}_j = -g^{pi}g_{ij} \mathcal{L}_{\rm matter} = -\delta^p_j \mathcal{L}_{\rm matter},
\eeq
which upon taking the trace and integrating becomes:
\beq
\frac{1}{V_3}\int d^3xdz\sqrt{-g}T^{j(2)}_j = -3 \int dz \sqrt{-g}\mathcal{L}_{\rm matter}= -3 L_g.
\eeq
The first term is instead more complicated, again requiring to rely on integration by parts and the equations of motion: differentiating with respect to $g^{ij}$ removes the abelian fields from the computation (with the homogeneous Ansatz only $\widehat{L}_{0z}$ is turned on), so we find
\beq
2\frac{\partial \mathcal{L}_{\rm matter}}{\partial g^{ij}} &=& -2M_5\left[ \Tr\left(L_{ik}L_{jk}\right)g^{kk}+\Tr \left(L_{iz}L_{jz}\right)g^{zz}  +\{L\leftrightarrow R \} \right]\nonumber\\
&=&-4a^{-2}(z)M_5\left[ \Tr \left(\frac{H^4}{4} \epsilon^{ika}\epsilon^{jkb}\tau^a\tau^b\right)\eta^{kk}+\Tr\left(\frac{H'^2}{4}\tau^i\tau^j \right)\eta^{zz}\right]\nonumber\\
&=&-a^{-2}(z)M_5\left[  2H^4\epsilon^{ika}\epsilon^{jka}\eta^{kk}+2H'^2 \delta^{ij}\eta^{zz}  \right].
\eeq
As before, we now raise one index, take the trace:
\beq
T^j_j=g^{ji}T^{(1)}_{ij} = -a^{-4}M_5\left[ 12H^4 +6H'^2 \right],
\eeq
and then integrate, using integration by parts:
\beq
\frac{1}{V_3}\int d^3xdz\sqrt{-g}T^{j(1)}_j &=& -\int dz\left[ 12M_5 a(z)H^4+6M_5 a(z)H'^2  \right]\\
&=&-M_5\int dz \left[12a(z)H^4-6\partial_z \left(a(z)\partial_z H\right) \right] H -6M_5\left[ a(z)H H' \right]^1_0\nonumber.
\eeq
As a first step, we recognize in the integral term the equation of motion for $H(z)$, so we plug it into our expression and then we integrate again by parts, to remove the derivative of the abelian field and make the presence of the Chern-Simons Lagrangian manifest:
\beq
\frac{1}{V_3}\int d^3xdz\sqrt{-g}T^{j(1)}_j &=& \int dz\left[ \frac{3N_c}{8\pi^2}\widehat{a}_0' H^3 \right] -6M_5\left[ a(z)HH' \right]^1_0\nonumber\\
&=&\frac{3N_c}{8\pi^2}\int dz \left[\partial_z\left( \widehat{a}_0 H^3\right)-3\widehat{a}_0 H^2H'\right]-6M_5\left[ a(z)HH' \right]^1_0\nonumber\\
&=&-3L_{CS} +\left[\frac{3N_c}{8\pi^2}\widehat{a}_0 H^3-6M_5 a(z)HH'\right]^1_0.
\eeq
We are left with the boundary term to evaluate: it is straightforward that the two terms evaluated on the UV boundary individually vanish, so that this is truly a purely infrared term. When evaluated on a solution of the equations of motion that is also in thermodynamic equilibrium, this boundary term is found to vanish (an evaluation we did numerically).
Putting all the results together, we can then write the following identifications, where with $L^\mathrm{on-shell}$ we mean the Lagrangian density integrated along $z$ (that is, the Lagrangian density from the point of view of the boundary theory):
\beq
\varepsilon=-L^\mathrm{on-shell}+\mu_B d\qquad ; \qquad P = L^\mathrm{on-shell},
\eeq
The holographic duality also identifies the grand canonical potential $\Omega$ with the on-shell Lagrangian, so that the usual thermodynamic relation for homogeneous matter
\beq
PV = -\Omega
\eeq
correctly emerges. The relation between energy density and pressure also is correctly reproduced:
\beq
\varepsilon = -P + \mu_B d.
\eeq
Note that to find all these thermodynamical results, we had to plug a thermodynamically stable configuration in the boundary term that enters the definition of $P$: this is however not unexpected, since thermodynamical and hydrodynamical pressure are in principle not the same quantity, but they are required to coincide at equilibrium. The stress-energy tensor computes the hydrodynamical pressure, hence the need to make use of the condition of thermodynamical equilibrium to connect with relations from thermodynamics.
\end{appendices}

\bibliographystyle{JHEP}
\bibliography{main}

\end{document}